\documentclass[aps,prd,notitlepage,twocolumn,superscriptaddress,showpacs]{revtex4-1}
\usepackage{amsfonts}
\usepackage{graphicx, bm, amsmath, amsfonts, amssymb}
\usepackage{tabularx}
\usepackage{caption}
\usepackage{subcaption}
\graphicspath{{figures/}}

\begin{document}
\title{Topological gravity in 3+1D and a possible origin of dark matter}
 
\author{Tianyao Fang}
 \affiliation{Department of Physics, The Chinese University of Hong Kong, Shatin, New Territories, Hong Kong, China}

\author{Zheng-Cheng Gu}
\email{zcgu@phy.cuhk.edu.hk}
 \affiliation{Department of Physics, The Chinese University of Hong Kong, Shatin, New Territories, Hong Kong, China}

\begin{abstract}
Dark matter is one of the deepest mystery of the universe. So far there is no natural explanation why the dark matter should exist and even dominate the universe. In this paper, we begin with a 3+1D topological gravity theory which is super renormalizable with vanishing beta functions, then we argue that Einstein gravity can emerge by condensing loop-like excitation from the underlying topological gravity theory. In the meanwhile, the uncondensed loop-like excitations serves as a natural candidate of dark matter and a generalized Einstein equation can be derived in the presence of loop-source(dark matter) background. Surprisingly, we find that such kind of dark matter will not contribute to scalar curvature, however, it will become a source of torsion. Finally, we derive the generalized Einstein equation in the presence of Dirac field. Very different from the usual Einstein-Carton theory, our theory further predicts that any type of normal matter, including Dirac field will not produce torsion. All these unique predictions can be tested by future experiments. Our framework suggests that topological invariant principle might play a more profound role than the well-known general covariance principle, especially towards understanding the nature of dark matter and quantum gravity in 3+1D. 
\end{abstract}

\maketitle

\section{Introduction}
\subsection{The dark matter puzzle}
It is well known that dark matter makes up about 27\% of the universe, which is much bigger than all normal matter - adds up to less than 5\% of the universe. Nevertheless, it is rather hard to observe dark matter directly due to their extremely weak interaction with normal matter. Actually it is long believed that dark matter only interacts with normal matter via gravity. 
On the other hand, Einstein's general relativity severs as the most profound framework to understand the basic structure of the universe, and the recent observation of gravitational wave by LIGO\cite{LIGO} further verifies the correctness of such an elegant theory. Therefore, it is very important to understand how dark matter can be included into Einstein equation in a natural and covariant way, and it is also very important to understand what makes dark matter so different from normal matter. Although many phenomenological models are proposed to describe dark matter\cite{Dark1,Dark2,Dark3}, none of them can be derived naturally from a general principle.  

\subsection{General covariance principle vs. topological invariance principle}
The general covariance principle, which demands a geometric, coordinate-independent formulation of physics, was long believed to be the mathematical foundation of Einstein's physical idea for general relativity. In particular, the Hilbert-Einstein action $S=\frac{1}{16\pi G}\int d^4x\sqrt{-g}R$ was first proposed by Hilbert based on such a novel principle to derive the well known (vacuum) Einstein equation. However, if we think about this principle more carefully, the covariance principle actually can not lead to Einstein equation in a unique way. For example, the higher order term $\sqrt{-g}R^2$ is also allowed by general covariance principle and such a term will induce higher order corrections to the Einstein equation. In usual quantum field theory such as gauge theory, one may argue that higher order terms should be suppressed by the renormalization group flow. Unfortunately, as a non-renormalizable theory, there is no natural reason to explain why higher order terms such as $\sqrt{-g}R^2$ should not appear in the semi-classical limit. Moreover, the general covariance principle is intrinsically inconsistent with quantum mechanics since the Schrodinger equation requires an absolute time.  

The topological aspects of Einstein gravity can be traced back to a century ago. The so-called Einstein-Carton action $S=\frac{1}{32\pi G }\int \varepsilon_{abcd}R^{ab}(\omega)\wedge e^{c}\wedge e^{d}$ was first proposed by Elie Cartan to describe spinors in curved spacetime where torsion was introduced. In such a first order formalism, Einstein gravity was reformulated as a topological field theory(TFT). 
Actually, the TFT framework not only provides us an alternative way to re-derive Einstein equation, but also has the unique advantage to eliminate higher order terms such as $R^2$.  In fact, the topological invariance principle might play a more profound role than the general covariance principle, and it might even help us to reveal the deep mystery of dark matter. 

In 2+1D, the Einstein-Carton action turns out to be exactly the same as the Chern-Simons action of Pioncare gauge group.\cite{CS1}
Three decades ago, Edward Witten proposed to use such a new framework to resovle the long standing renormalizabilty puzzle of 2+1D quantum gravity(at least perturbatively).
Although $2+1$D gravity is somewhat trivial due to the absence of propagating gravitational wave and vanishing of space-time curvature, it still provides us a concrete example of understanding quantum gravity via topological quantum field theory(TQFT). Moreover, according to the correspondence between Chern-Simons theory and CFT, the ADS3/CFT2 correspondence conjecture can also be understood in a very natural way\cite{ADSCFT3}.

Nevertheless, the TQFT approach can not be easily generalized into $3+1$D due to the following difficulties. (a) Einstein's gravity in $3+1$D contains propagating mode -- the gravitational wave, therefore it is obviously not a TQFT in the usual sense. (b) Our knowledge of higher dimensional TQFT is very limited and there is no Chern-Simons like action in $3+1$D. Thanks to the recent development of the classification of topological phases of quantum matter in higher dimensions\cite{Chenlong,Wenscience,cobordism,Wencoho}, new types of TQFT have been discovered in $3+1$D to describe the so-called three-loop-braiding statistics. Here we argue that such types of TQFT are closely related to Einstein gravity. In particular, we conjecture that gravitational wave will disappear at an extremely high energy scale and $3+1$D quantum gravity might be controlled by a TQFT renormalization group fixed point.  


\subsection{Main results and outline}
In this paper, we first review a topological quantum gravity theory in 3+1D, which can be regarded as a higher dimensional generalization of Chern-Simons action of Pioncare gauge group. In the semi-classical limit, we argue that the condensation of loop-like excitations in 3+1D topological quantum gravity will lead to Einstein equation in a natural way. We conjecture that uncondensed loop-like exciations can be a very promising candidate of dark matter. By adding the loop source term, we can derive a generalized Einstein equation which naturally includes dark matter sector. Surprisingly, we find that such kind of dark matter will not contribute to scalar curvature and this unique property makes the dark matter very different from the (massive) normal matter. Finally, we derive the generalized Einstein equation in the presence of Dirac field. Very different from the usual Einstein-Carton theory, we predict that loop current is the only source of torsion while any type of normal matter, including Dirac field will not contribute to torsion. All these unique predictions can be tested in future experimental measurement. 

The rest of the paper is organized as follows. In section \ref{topgravity}, we will review the topological gravity theory in 3+1D and show that such a theory is actually super renormalizable with vanishing beta functions. Then we will study the canonical quantization of this theory in details. In section \ref{loopcondense}, we will show how Einstein gravity might emerge from the underlying topological gravity theory in 3+1D by adding a topological mass term of the 2-form gauge field. Physically, we conjecture that such a phenomenological theory might describe a "loop condensing" phase.
Remarkably, we find that the whole theory is still super renormalizable with vanishing beta functions. In section \ref{loopsource}, we propose the loop-like extensive object as a natural candidate of dark matter and derive a generalized Einstein equation in the presence of loop source term.  In section \ref{Dirac}, we also study the generalized Einstein equation in the presence of Dirac fields. In section \ref{general}, we study the most general form of topological action and discuss the uniqueness of the generalized Einstein equation. Finally, there is a conclusion and discussion section. 

\section{Topological gravity in 3+1D}
\label{topgravity}
Let us begin with the topological gravity theory in 3+1D\cite{Topgravity}. Consider the following topological invariant action:
\begin{eqnarray}
S_{top}&=&\frac{k_{1}}{4\pi }\int \varepsilon
_{abcd} R^{ab}\wedge e^{c}\wedge e^{d} +\frac{k_{2}}{2\pi }\int B_{ab} \wedge  R^{ab} \nonumber\\ &&+\frac{k_{3}}{2\pi }%
\int \widetilde{B}_{a} \wedge  T^{a}, \label{action}
\end{eqnarray}
where $e$ is the tetrad field, $R$ is the curvature tensor, $T$ is the torsion tensor and $B,\widetilde{B}$ are 2-form gauge fields. It is straightforward to verify that the above action is invariant under the following (twisted) 1-form and 2-form gauge transformations, respectively:
\begin{eqnarray}
e^{a} &\rightarrow &e^{a}+Df^{a} \nonumber\\ 
B_{ab} &\rightarrow &B_{ab}-\frac{k_{3}}{2k_{2}}\left( \widetilde{B}%
_{a}f_{b}-\widetilde{B}_{b}f_{a}\right)  \nonumber \\
\widetilde{B}_{a} &\rightarrow &\widetilde{B}_{a}-\frac{k_{1}}{k_{3}}
\varepsilon _{abcd}f^{b}R^{cd}, \label{1form}
\end{eqnarray}
and
\begin{eqnarray}
B_{ab} & \rightarrow & B_{ab}+D\xi _{ab}, \label{2forma}\\
\widetilde{B}_{a} &\rightarrow &\widetilde{B}_{a}+D \tilde{\xi} _{a} 
\nonumber\\
B_{ab} &\rightarrow &B_{ab}-\frac{k_{3}}{2k_{2}}\left( \tilde{\xi}  _{a}\wedge
e_{b}-\tilde{\xi}  _{b}\wedge e_{a}\right).  \label{2formb}
\end{eqnarray}
Such an action can be regarded as the non-Abelian generalization of $AAdA+BF$ type TQFT\cite{aada1,aada2,aada3} of Poincare gauge group. Physically, it has been shown that such kind of TQFT describes the so-called three-loop-braiding statistics\cite{loop1,loop2}. Apparently, as a TQFT, the action Eq. (\ref{action}) is a super renormalizable theory. Below we will discuss the coefficient quantization and canonical quantization of such a theory. 

\subsection{Coefficient quantization and super renormalizability}
Since the action Eq. (\ref{action}) contains a cubic term, it is not easy to perform the usual momentum shell path integral scheme to understand its renormalizability. Similar to the Chern-Siomns theory in 2+1D, below we will show that all the coefficients $k_1,k_2$ and $k_3$ are quantized, which implies the vanishing of beta functions and super renormalizability. 

We first consider the coefficient quantization of the $BR$ term. We note that the 2-form gauge transformation Eq. (\ref{2forma}) will induce a boundary term for $S_{top}$.

\begin{eqnarray}
\delta S_{top} &=&\frac{k_{2}}{2\pi }\int D\xi _{ab}\wedge R^{ab} \nonumber\\
\ &=&\frac{k_{2}}{2\pi }\int \left[ D\left( \xi _{ab}\wedge
R^{ab}\right) +\xi _{ab}\wedge DR^{ab}\right] \nonumber\\
&=&\frac{k_{2}}{2\pi }\int D\left( \xi _{ab}\wedge R^{ab}\right) \nonumber\\
&=&\frac{k_{2}}{2\pi }\int d\left( \xi _{ab}\wedge R^{ab}\right),
\end{eqnarray}
where we use the identity $DR=0$ in the second line(see Appendix \ref{structure}), and in the last equation we use the fact that the covariant differentiation on a scalar
is just the same as usual partial differentiation. Considering a manifold with a boundary, such a term becomes:
\begin{equation}
\delta S_{top}=\frac{k_{2}}{4\pi }\int \xi _{ab}\wedge R^{ab}. 
\end{equation}

To understand the quantization of $k_2$, we just need to consider the maximum compact subgroup of $SO(3,1)$, which is $SO(3)$. If we choose $\xi _{ab}=\varepsilon _{abc}e^{c}$ where $e$ is a tetrad field of 3D Euclidean space, the boundary term can be simplified as:
\begin{equation}
\delta S_{top}^\prime=\frac{k_{2}}{4\pi }\int \varepsilon _{abc}e^{a}\wedge R^{bc} ,
\end{equation}
which is exactly the Chern-Simons action of gauge field $ISO(3)$ with quantized coefficient $k_2=2\mathbb{Z}$. (See Appendix \ref{CS} for full details.) 
 
Next we consider the coefficient quantization of $Ree$ term.  
Under the gauge transformation Eq. (\ref{1form}), the torsion tensor $T$ will be transformed as:
\begin{equation}
T^{a}\rightarrow T^{a}+R^{ab}f_{b}.
\end{equation}
Up to linear order, we obtain the variation of the action $S_{top}$ as:
\begin{eqnarray}
\delta S_{top} &=&\frac{k_{1}}{2\pi }\int \varepsilon _{abcd} R^{ab}\wedge
Df^{c}\wedge e^{d}-\frac{k_{2}}{2\pi }\frac{k_{3}}{k_{2}}\int \widetilde{B}_{a}f_{b} \wedge R^{ab}\nonumber\\
&&+\frac{k_{3}}{2\pi }\int \widetilde{B}_{a}\wedge R^{ab}f_{b}-\frac{k_{3}}{2\pi }\frac{k_{1}}{k_{3}}\int \varepsilon _{abcd}T^{a}f^{b}\wedge R^{cd} 
\nonumber \\
&=&\frac{k_{1}}{2\pi }\int \varepsilon _{abcd}\left( Df^{a}\wedge
e^{b}-T^{a}f^{b}\right) \wedge R^{cd}  \nonumber \\
&=&\frac{k_{1}}{2\pi }\int \varepsilon _{abcd}\left( Df^{a}\wedge
e^{b}+f^{a}De^{b}\right) \wedge R^{cd}  \nonumber \\
&=&\frac{k_{1}}{2\pi }\int \varepsilon _{abcd}d\left( f^{a}e^{b}\wedge
R^{cd}\right), \label{k1}
\end{eqnarray}
where we derive the last line by using $DR=0$. It is obvious that if we choose $f^{0}=1$ and $f^{i}=0$,  Eq. (\ref{k1}) will have exactly the same boundary term as $\delta S_{top}^\prime$. Therefore, we conclude that $k_{1}$ is also quantized as $k_{1}=2\mathbb{Z}$.
 
Finally we consider the quantization of $k_{3}$. Under the second 2-form gauge transformation in Eq. (\ref{2formb}), the variation of total action reads:

\begin{eqnarray}
\delta S_{top} &=&-\frac{k_{2}}{2\pi }\frac{k_{3}}{k_{2}}\int \tilde{\xi}
_{a}\wedge e_{b}\wedge R^{ab}+\frac{k_{3}}{2\pi }\int T^{a}\wedge
D\tilde{\xi} _{a} \nonumber\\
&=&-\frac{k_{3}}{2\pi }\int \tilde{\xi} _{a}\wedge e_{b}\wedge R^{ab}+\frac{%
k_{3}}{2\pi }\int D\left( e^{a}\wedge D\tilde{\xi} _{a}\right)\nonumber\\ &&+\frac{k_{3}}{%
2\pi }\int e^{a}\wedge R_{a}^{\ b}\wedge \tilde{\xi} _{b}  \nonumber \\
&=&\frac{k_{3}}{2\pi }\int D\left( e^{a}\wedge D\tilde{\xi} _{a}\right) .
\end{eqnarray}
If we choose $\tilde{\xi} ^{a}=e^{a}$, and consider it on a flat space-time,
that is, $\omega _{\ b}^{a}=R_{\ b}^{a}=0,$ the equation above becomes:

\begin{equation}
\delta S_{top}=\frac{k_{3}}{2\pi }\int de^{a}\wedge de_{a}=\frac{k_{3}}{%
2\pi }\int F^{a}\wedge F_{a} ,
\end{equation}
where the field strength of $e^{a}$ is defined as $F^{a}=\frac{1}{2}\left(
\partial _{\mu }e_{\nu }^{a}-\partial _{\nu }e_{\mu }^{a}\right) dx^{\mu
}\wedge dx^{\nu }$. We note that the first Chern class $\frac{1}{\left( 2\pi
\right) ^{2}}\int F^{a}\wedge F_{a}=c_{1}$ should be an integer\footnote{As a TQFT, here we define the tetrad field $e$ as a compact $U(1)$ gauge field}, thus we finally have:

\begin{equation}
\delta S_{top}=2\pi k_{3}c_{1}.
\end{equation}
Again, as the variation of the total action should be some integer multiplied by $2\pi $, we conclude that $k_{3}=\mathbb{Z}$.

Apparently, as a TQFT, the action Eq. (\ref{1form}) is a super renormalizable quantum theory with vanishing beta functions due to the coefficient qauntizations. 
It is easy to see that $e$ and $\omega$ are of dimension 1, $B$ and $\widetilde{B}$ are of dimension 2. 
Thus, when integrating out the fast mode $\Lambda/s<p<\Lambda$ and recovering the integral region by doing the rescaling transformation $p^{\prime}=sp$, we will have the following rescaling law of the fields: 

\begin{eqnarray}
e^{\prime }\left( p^{\prime }\right) &=&s^{-3}e_{<}\left( p^{\prime
}/s\right)  \nonumber\\
\omega^{\prime }\left( p^{\prime }\right) &=&s^{-3}\omega_{<}\left( p^{\prime}/s\right)\nonumber\\
B^{\prime }\left( p^{\prime }\right) &=&s^{-2}B_{<}\left( p^{\prime
}/s\right)  \nonumber\\
\widetilde{B}^{\prime }\left( p^{\prime }\right) &=&s^{-2}\widetilde{B}_{<}\left( p^{\prime}/s\right).  
 \label{rescal}
\end{eqnarray}

\subsection{Canonical quantization} 
Now we consider the conanical quantization of the 3+1D topological gravity theory. Without loss of generality, here and below we will choose $k_1=k_2=2$ and $k_3=1$ for convenient. The canonical momentums of $\omega _{i}^{ab}$ and $e_{i}^{a}$ can be defined as $\Pi _{ab}^{i}=\frac{1}{2\pi}\varepsilon
^{ijk}\varepsilon _{abcd}e_{j}^{c}e_{k}^{d}+\frac{1}{2\pi}\varepsilon
^{ijk}B_{abjk}$ and $\pi _{a}^{i}=\frac{1}{4\pi}\varepsilon ^{ijk}\widetilde{B}_{ajk}$. The Lagrangian density of Eq. (\ref{1form}) can be rewritten as:
\begin{widetext}
\begin{eqnarray}
\mathcal{L}_{Top} &=&\Pi _{ab}^{i}\partial _{0}\omega _{i}^{ab}+\pi _{a}^{i}\partial _{0}e_{i}^{a}+\frac{1}{2\pi}\varepsilon
^{ijk}B_{ab0i}R_{jk}^{ab}+\frac{1}{4\pi}\varepsilon ^{ijk}\widetilde{B}
_{a0i}T_{jk}^{a}+\frac{1}{\pi} e_{0}^{a}(D_{i}\pi _{a}^{i}+\frac{1}{2}\varepsilon^{ijk}\varepsilon _{abcd}R_{ij}^{bc}e_{k}^{d})\nonumber\\&&+ \omega _{0}^{ab}(D_{i}\Pi _{ab}^{i}+\frac{1}{2}(\pi _{a}^{i}e_{bi}-\pi _{b}^{i}e_{ai})) .
\end{eqnarray}
\end{widetext}

Canonical quantization requires:
\begin{eqnarray}
\lbrack \omega _{i}^{ab}(x),\Pi _{cd}^{j}(y)]=i\delta
_{i}^{j}\delta _{cd}^{ab}\delta (x-y), \nonumber\\
\lbrack e_{i}^{a}(x),\pi _{b}^{j}(y)]=i\delta _{i}^{j}\delta
_{b}^{a}\delta (x-y), \nonumber\\
\ all\  other\  commutators=0,  
\end{eqnarray}
which are equivalent to:
\begin{eqnarray}
\lbrack \omega _{i}^{ab}(x),B_{cdjk}(y)]&=&i\pi \varepsilon _{ijk}\delta
_{cd}^{ab}\delta (x-y), \nonumber\\ 
\lbrack e_{i}^{a}(x),\widetilde{B}_{bjk}(y)]&=&i2\pi\varepsilon _{ijk}\delta
_{b}^{a}\delta (x-y), \nonumber\\ 
\lbrack B_{abij}(x),\pi _{c}^{k}(y)]&=&i\varepsilon _{abcd}(e_{i}^{d}(x)\delta _{j}^{k}-e_{j}^{d}(x)\delta _{i}^{k})\delta (x-y),\nonumber	\\
\end{eqnarray}
with the following flat-connection constraints:
\begin{eqnarray}
\mathcal{A}^{abi}&=&\frac{1}{2}\varepsilon ^{ijk}R_{jk}^{ab}=0,\nonumber\\ 
\mathcal{B}^{ai}&=&\frac{1}{2}\varepsilon ^{ijk}T_{jk}^{a}=0, \nonumber\\ 
\mathcal{C}_{a}&=&D_{i}\pi _{a}^{i}+\frac{1}{2}\varepsilon ^{ijk}\varepsilon
_{abcd}R_{ij}^{bc}e_{k}^{d}=0, \nonumber\\ 
\mathcal{D}_{ab}&=&D_{i}\Pi _{ab}^{i}+\frac{1}{2}(\pi _{a}^{i}e_{bi}-\pi
_{b}^{i}e_{ai})=0.
\end{eqnarray}

It can be shown that these constraints are closed and satisfy the following commutation relations:
\begin{eqnarray} 
\lbrack \mathcal{A}^{abi}(x),\mathcal{D}_{cd}(y)]&=&-2i\delta _{[c}^{[a} \mathcal{A}_{\ d]}^{b]\ i}(x)\delta (x-y),
\nonumber\\ 
\lbrack \mathcal{B}^{ai}(x),\mathcal{C}_{b}(y)]&=&-i \mathcal{A}_{\ b}^{a\ i}(x)\delta (x-y) ,
\nonumber\\
\lbrack \mathcal{B}^{ai}(x),\mathcal{D}_{cd}(y)]&=&i\delta _{[c}^{a} \mathcal{B}_{d]}^{i}(x)\delta (x-y),
\nonumber\\
\lbrack \mathcal{C}^{a}(x),\mathcal{D}_{cd}(y)]&=&i\delta _{[c}^{a} \mathcal{C}_{d]}(x)\delta (x-y)\nonumber\\
\lbrack \mathcal{D}^{ab}(x),\mathcal{D}_{cd}(y)] &=& 4i\delta _{[c}^{[a} \mathcal{D}_{\ d]}^{b]}(x)\delta (x-y),\nonumber\\ 
\rm{others}&=&0
\end{eqnarray}
The proof of the above relations involves very tedious calculations and we will leave the full details in Appendix \ref{constraints}.

Similar to the 2 + 1D topological gravity, the phasespace
to be quantized is exactly the solutions of the above
constraints divided by the group of gauge transformations
generated by the constraints. The quantum Hilbert
space is the 
flat connections of Poincare group modulo
gauge transformations.(If we regard $e$ and $\omega$ as coordinates
while $\Pi$ and $\pi$ as momentums.)

Finally, we would like to stress that just like we use tensor category theory to describe Chern-Simons theory, the algebraic tensor 2-category theory\cite{TCAT1,TCAT1} might also provide us an equivalent UV-complete description for such a topological quantum gravity theory in $3+1$D.  

\section{Loop condensation and the emergence of vacuum Einstein equation}
\label{loopcondense}
To this point, one may wonder why we are interested in the $3+1$D topological gravity theory which is somewhat trivial. Here we conjecture that quantum gravity is actually controlled by a topological gravity fixed point and the classical space-time vanishes at extremely high energy scale. Therefore it is quite natural to expect the vanishing of curvature and torsion at that scale. 
A possible way to generate interesting dynamics is condensing loop-like excitations(i.e., flux lines in the context of gauge theory) of a $3+1$D TQFT.
If we further assume that the condensed loops carry a nontrivial linking Berry phase\cite{BB1,BB2,BB3}, a $\int Tr(B\wedge B)$ type topological mass term can be generated. Let us consider the following action:
\begin{eqnarray}
S_\theta=-\frac{\theta}{2\pi} \int B_{ab}\wedge B^{ab}.
\end{eqnarray}
Apparently, such a term breaks the 2-form gauge symmetry as well as the translational gauge symmetry explicitly. A microscopic derivation of the above term from loop condensation is beyond the scope of this paper. Here we just introduce such a term phenomenologically to describe low energy dynamics and ignore all the microscopic details of loop dynamics, which is the analog of using massive gauge boson to describe Abelian Higgs phase and considering the infinite massive limit for Higgs boson. 

The classical equation of motion(EOM) for the total action $S$ reads:
\begin{eqnarray}
&& B^{ab}=\frac{1}{\theta}R^{ab},\quad T^a=0, \quad 
\varepsilon_{abcd} e^b \wedge R^{cd}= -\nabla \tilde{B}_a, \nonumber\\ &&\varepsilon_{abcd} T^c\wedge e^d+\frac{1}{2}(\tilde{B}_a \wedge e_b-\tilde{B}_b \wedge e_a)=-\nabla B_{ab}.
\end{eqnarray}
Insert the first two equations into the last equation, we have:
\begin{eqnarray} 
\frac{1}{2}(\tilde{B}_a \wedge e_b-\tilde{B}_b \wedge e_a)=-\frac{1}{\theta}\nabla R_{ab}=0.
\end{eqnarray}
The equation above can be rewritten in a compact form as $\varepsilon^{abcd} \tilde{B}_a\wedge e_b=0$, which further implies $\tilde{B}^a=0$. Thus, we eventually derive the vacuum Einstein-Cartan equation: 
\begin{eqnarray}
\varepsilon_{abcd} e^b \wedge R^{cd}=0.
\end{eqnarray}
Remarkably, for small $\theta$, the total action $S=S_{\rm{Top}}+S_\theta$ is still power-counting renormalizable since $S_\theta$ only contains dimension 4 operator. Below we will show that the beta function still vanishes even in the presence of $S_\theta$ term. 

Since $S_{\theta }$ is a quadratic term, there is no mixed mode for this term. Thus, we can separate the total action as:
\begin{equation}
S=S_{0<}+S_{0>}+S_{0<,>}+S_{\theta<}+S_{\theta>} ,\label{sep}
\end{equation}
where $S_0 \equiv S_{top}$ is the unperturbed action. To simplify the notation, here $<$ represents slow mode, $>$ represents fast mode and $<,>$ represents mixed mode.

Thus, the partition function can be rewritten as:
\begin{widetext}
\begin{eqnarray}
Z &=& \int D_{<} e^{i(S_{0<}+S_{\theta <})} \int D_{>} e^{i(S_{0>}+S_{0<,>}+S_{\theta >})}  \nonumber \\
&=& \int D_{<} e^{i(S_{0<}+S_{\theta <})} \frac{\int D_{>} e^{iS_{\theta>}} e^{i(S_{0>}+S_{0<,>})}}{\int D_{>} e^{i(S_{0>}+S_{0<,>})}}\int D_{>} e^{i(S_{0>}+S_{0<,>})} \nonumber \\
&=& \int D_{<} e^{i(S_{0<}+S_{\theta<})} \langle e^{iS_{\theta>}}  \rangle_{0}, \label{Z}
\end{eqnarray}
where $D_{<}(D_{>})$ is the integral over the slow (fast) modes of $e$, $\omega$, $B$ and $\widetilde{B}$. Here $\langle\rangle_{0}$ denotes averaging with respect to $S_{0>}+S_{0<,>}$. Since the beta function of $S_{0}$ vanishes, $\int D_{>} e^{i(S_{0>}+S_{0<,>})}$ must be a constant which is independent of slow modes, thus we can drop it in the last line. 

Let us denote $e^{i\delta S}=\langle e^{iS_{\theta>}}  \rangle_{0}$ and we will show such a term actually vanishes up to any order. To illustrate the basic idea for the proof, below we just expand $\delta S$ up to the second order by using cumulant expansion:
\begin{equation}
i\delta S \simeq i\langle S_{\theta>}\rangle_{0}- \frac{1}{2}(\langle S_{\theta>}^{2}\rangle_{0}-\langle S_{\theta>}\rangle_{0}^{2} ).
\end{equation}
We first consider the term $i\langle S_{\theta>}\rangle_{0}$, which can be computed from the following generating functional:
\begin{eqnarray}
Z[J]=\int D_{>} \exp \left[i(S_{0>}+S_{0<,>})+ \frac{i}{\pi}\int dp J^{ab}(p)\wedge B_{>ab}(-p) \right]. \nonumber\\
\end{eqnarray}
The expectation values can be obtained by differentiation of the generating functional with respect to the source $J$:
\begin{equation}
\langle S_{\theta>}\rangle_{0}=\frac{\pi\theta}{32} Z[0]^{-1}\int dp\varepsilon_{\mu\nu\rho\sigma}(-i\frac{\delta}{\delta J(p)_{ab\mu\nu}})(-i\frac{\delta}{\delta J(-p)^{ab}_{\rho\sigma}})Z[J]\mid_{J=0},
\end{equation}
As $ S_{0>}+S_{0<,>}$ depends on $B_{>}$ linearly, performing the integration $\int DB_{>}$ will lead to delta functions. All terms containing $B_{>}$ can be expressed as:
\begin{eqnarray}
S_{B_{>}}&=&\frac{1}{\pi}\int dp (J^{ab}(p)+d\widetilde{\omega}^{ab}(p))\wedge B_{>ab}(-p)+\frac{1}{\pi}\int dp_{1} dp_{2}dp_{3}[\Omega^{ac}(p_{1})\wedge\Omega^{\ b}_{c}(p_{2}) \nonumber\\
&&+2\widetilde{\omega}^{ac}(p_{1})\wedge\Omega^{\ b}_{c}(p_{2})+\widetilde{\omega}^{ac}(p_{1})\wedge\widetilde{\omega}^{\ b}_{c}(p_{2})]\wedge B_{>ab}(p_{3})\delta(p_{1}+p_{2}+p_{3}). 
\end{eqnarray}
Here we denote
$\omega_{>}^{ab}\equiv \widetilde{\omega}^{ab}$ and $\omega_{<}^{ab}\equiv\Omega^{ab}$ for convenience, and we also use $\int dp$ as a simplified notation of $\int \frac{d^4 p}{(2\pi)^4}$.
When we carry out functional integral of the fast mode, we regard all slow modes as fixed parameters. After integrating over $B_{>}$, the generating functional becomes:
\begin{equation}
Z[J]=\int De_>D\widetilde{B}_>\int \prod_{p}D\widetilde{\omega}(p)\delta(f_{ab\mu\nu}(p))e^{iS'}, \label{partition}
\end{equation}
where $S'$ is the action that does not contain $B_{>}$ and $f$ is a function of $\widetilde{\omega},\Omega,J$ in the form:
\begin{eqnarray}
f_{ab\mu\nu}(p)
&=& \frac{1}{\pi}[\frac{1}{2} J_{ab\mu\nu}(p)-ip_{\mu}\widetilde{\omega}_{ab\nu}(p)+\int dp'[\Omega_{ac\mu}(p')\Omega^{c}_{\ b\nu}(p-p')+\widetilde{\omega}_{ac\mu}(p')\Omega^{c}_{\ b\nu}(p-p')-\widetilde{\omega}_{bc\mu}(p')\Omega^{c}_{\ a\nu}(p-p')
\nonumber\\
&&+\widetilde{\omega}_{ac\mu}(p')\widetilde{\omega}^{c}_{\ b\nu}(p-p')]]-\mu\leftrightarrow\nu  .\label{fomega}
\end{eqnarray}

Now we need to solve $f(p)_{ab\mu\nu}=0$ for $\widetilde{\omega}_{ab\mu}(p)$ at each momentum. However, for a fixed momentum $p$, there are 36 equations but only 24 unknown $\widetilde{\omega}_{ab\mu}(p)$.  
In general, such kind of overdetermined system has no solution and $Z[J]$ must be zero. Below we give a more rigorous proof for $Z[J]=0$ for small but finite $J$.

Since $Z[0]$ is nonzero due to the renormalizable property of $S_{top}$, Eq. (\ref{fomega}) must have a solution for arbitrary $\Omega_{ab\mu}(p)$ at each $p$ in momentum shell when $J_{ab\mu\nu}=0$. For small but finite $J$, we assume the solutions of $\widetilde{\omega}$ can be expanded around the solutions for $J=0$
with $\widetilde{\omega}=\widetilde{\omega}_{0}+\delta\widetilde{\omega}$ and $\widetilde{\omega}_{0}$ satisfies:
\begin{eqnarray}
&&-ip_{\mu}\widetilde{\omega}_{0ab\nu}(p)+\int dp'[\Omega_{ac\mu}(p')\Omega^{c}_{\ b\nu}(p-p')+\widetilde{\omega}_{0ac\mu}(p')\Omega^{c}_{\ b\nu}(p-p') -\widetilde{\omega}_{0bc\mu}(p')\Omega^{c}_{\ a\nu}(p-p')\nonumber\\
 &&+\widetilde{\omega}_{0ac\mu}(p')\widetilde{\omega}^{\ c}_{0\ b\nu}(p-p')]-\mu\leftrightarrow\nu=0. \label{j=0}
\end{eqnarray}
We begin with a simple case with $\Omega=0$. Eq. (\ref{j=0}) becomes:
\begin{equation}
-ip_{\mu}\widetilde{\omega}_{0ab\nu}(p)+\int dp'\widetilde{\omega}_{0ac\mu}(p')\widetilde{\omega}^{\ c}_{0\ b\nu}(p-p')-\mu\leftrightarrow\nu=0.
\end{equation}
Obviously $\widetilde{\omega}_{0}=0$ is a solution of the above equation. Up to first order of $\delta\widetilde{\omega}$, Eq. (\ref{fomega}) can be linearized as:
\begin{equation}
ip_{\mu}\delta\widetilde{\omega}_{ab\nu}(p)-ip_{\nu}\delta\widetilde{\omega}_{ab\mu}(p)=J_{ab\mu\nu}(p).
\end{equation}
By multiplying $\varepsilon^{\mu\nu\rho\sigma}p_{\rho}$ on both sides, we finally obtain:
\begin{equation}
\varepsilon^{\mu\nu\rho\sigma}J_{ab\mu\nu}(p)p_{\rho}=0,
\end{equation}
which has no solutions for general source term $J_{ab\mu\nu}(p)$.

Now we consider general $\Omega$. For small but finite $J$, we can again only keep up to first of $\delta\widetilde{\omega}$ for Eq. (\ref{fomega}). Again, by multiplying $\varepsilon^{\mu\nu\rho\sigma}$ on both sides, we obtain:
\begin{equation}
\varepsilon^{\mu\nu\rho\sigma}\lbrace-ip_{\mu}\delta\widetilde{\omega}_{ab\nu}(p)+\int dp'[\omega'_{ac\mu}(p')\delta\widetilde{\omega}^{c}_{\ b\nu}(p-p')-\omega'_{bc\mu}(p')\delta\widetilde{\omega}^{c}_{\ a\nu}(p-p')]\rbrace=-\frac{1}{2} \varepsilon^{\mu\nu\rho\sigma}J_{ab\mu\nu}(p), \label{omeganon}
\end{equation}
where $\omega'_{ab\mu}=\Omega_{ab\mu}+\widetilde{\omega}_{0ab\mu}$. In general, such an overdetermined nonhomogeneous linear equation system also has no solution for general source term $J$. 

\end{widetext}


Thus, we conclude that the generating function $Z[J]$ must be zero for at least small but finite source term $J$, and all the following expectation values must vanish.
\begin{equation}
\langle S_{\theta>}\rangle_{0}=\langle S^{2}_{\theta>}\rangle_{0}=...=0.
\end{equation}
This is simply because all the higher order terms can also be derived by applying derivatives with respect to the generating functional $Z[J]$. Therefore, we have $\delta S=0$ or $\langle e^{iS_{\theta>}}\rangle_{0}=1$. On the other hand, the slow mode part reads:
\begin{equation}
S_{\theta <} =-\frac{\theta }{8\pi }\int dp_{1}\varepsilon^{\mu\nu\rho\sigma}B_{<\mu\nu}^{\ ab}\left( p_{1}\right)
B_{< ab\rho\sigma}\left( -p_{1}\right).
\end{equation}
After recovering the integral region and rescaling of fields, $S_{\theta }$ becomes:
\begin{eqnarray}
S_{\theta }^{\prime }&=&-s^{-4+2+2}\frac{\theta }{8\pi }\int dp'_{1}\varepsilon^{\mu\nu\rho\sigma}B^{\prime ab}_{\mu\nu}\left(p_{1}^{\prime }\right) B^{\prime }_{ ab\rho\sigma}\left( -p_{1}^{\prime }\right) \nonumber\\
&=&\frac{\theta }{8\pi }\int dp'_{1}\varepsilon^{\mu\nu\rho\sigma}B^{\prime ab}_{\mu\nu}\left(p_{1}^{\prime }\right) B^{\prime }_{ ab\rho\sigma}\left( -p_{1}^{\prime }\right),
\end{eqnarray}
which has exactly the same form as $S_{\theta }$. Thus the beta function of $S_\theta$ term vanishes as well, and the whole theory is still super renormalizable.

Finally, we would like to stress that although the total action $S$ is super renormalizable, it does not imply a UV-complete quantum gravity theory due to the explicit broken of 2-form gauge symmetries by the $S_\theta$ term. How to develop a theoretical framework, e.g. topological string field theory to describe the loop condensation phase without 2-form gauge symmetry breaking could be an essential step towards establishing a UV-complete theory of quantum gravity, and we will leave this difficulty problem in our future study. Below we will discuss the semi-classical consequence of loop condensation and discuss it's potential connection with the origin of dark matter.

\section{Generalized Einstein equation and a potential origin of dark matter}
\label{loopsource}
In the previous section, we argue that the condensation of loop-like excitations coupled with the 2-form gauge field $B$(more precisely, that is the flux line of gauge field $\omega$, and we will call it $\omega$-loop below) will naturally lead to the emergence of vacuum Einstein equation. However, those loop-like excitations coupled with 2-form gauge field $\widetilde{B}$(more precisely, that is the flux line of $e$ gauge field, and we will call it $e$-loop below) could not condense simultaneously due to the presence of $Ree$ term which describes non-trivial three-loop braiding process between $\omega$-loop and $e$-loop with $e$-loop as the base loop. Therefore, linked $e$-loops must be confined and they can not be observed at low energy. Nevertheless, those unlinked $e$-loop still can survive at low energy and these extensive objects could be a natural candidate of dark matter and the framework developed here will allow us to derive a generalized Einstein equation with loop-like extensive matter source.

Since the $B\wedge B$ term also breaks the 1-form gauge transformation Eq. (\ref{1form}), the expectation value of $e$ could also be non-zero in the loop condensation phase:
\begin{equation}
\left\langle e_{\mu }^{a}\right\rangle =\frac{1}{l_{p}}\delta _{\mu }^{a} ,
\end{equation}
where $l_p$ is a fundamental length scale induced by loop condensation, and we can naturally interpret such a length scale as the Planck length, which uniquely determines the Newton's constant. Moreover, the non-zero expectation value of $e$ also naturally explains the origin of a flat background metric of the universe. In our usual convention, the flat background metric is defined as $e_{\mu }^{a}=\delta _{\mu }^{a}$, thus it is more convenient to redefine $\widetilde{e}=l_{p}e$ such that $\left\langle \widetilde{e}_{\mu }^{a}\right\rangle =\delta _{\mu }^{a}$. The $S_{top}$ can be rewritten as:(Without confusing, we just replace all $\widetilde{e}$ by $e$)
\begin{eqnarray}
S_{top}&=&\frac{1}{2\pi l_{p}^{2}}\int R^{ab}\wedge e^{c}\wedge e^{d}\varepsilon
_{abcd}+\frac{1}{\pi}\int R^{ab}\wedge B_{ab} \nonumber\\ &&+\frac{1}{2\pi l_{p}}\int T^{a}\wedge 
\widetilde{B}_{a}.
\end{eqnarray}

To simplify the coefficients in our future calculation, we further rescale $B\rightarrow \pi B$, $\widetilde{B}\rightarrow 2\pi\widetilde{B}$, and define $\pi l_{p}^{2}=\kappa=16\pi G$, where $G$ is the Newton constant. Thus, $S_{top}$ can be rewritten as:
\begin{eqnarray}
S_{top}&=&\frac{1}{2\kappa}\int R^{ab}\wedge e^{c}\wedge e^{d}\varepsilon
_{abcd}+\int R^{ab}\wedge B_{ab} \nonumber\\ &&+\frac{1}{l_{p}}\int T^{a}\wedge 
\widetilde{B}_{a}. \label{S_top}
\end{eqnarray}

Now, we consider the total action $S$ with both $S_\theta$ and loop-source term coupled with the 2-form gauge field $\widetilde{B}$:
\begin{equation}
S=S_{top}-\frac{\theta\pi}{2} \int B^{ab}\wedge B_{ab}-\int \sigma ^{a}\wedge 
\widetilde{B}_{a},
\end{equation}
where $\sigma ^{a}$ is the $e$-loop current and $\sigma ^{a}\wedge \widetilde{B}_{a}$ is a topological invariant (minimal) coupling term.

Below we will derive the generalized Einstein equation from the above total action $S$.

(1)Variation with respect to $B$ gives:
\begin{eqnarray}
R^{ab}=\theta \pi B^{ab},\  \text{in terms of components}\  R_{\mu \nu }^{ab}=\theta \pi B_{\mu \nu
}^{ab}.\nonumber \\ \label{RBrelation}
\end{eqnarray}

(2)Variation with respect to $\widetilde{B}$ gives:
\begin{equation}
T^{a}=l_{p}\sigma ^{a},\  \text{in terms of components}\  T_{\mu \nu }^{a}=l_{p}\sigma
_{\mu \nu }^{a}.
\end{equation}

(3)Variation with respect to $\omega $ gives:
\begin{widetext}
\begin{equation}
\frac{1}{2\kappa}\int D\delta \omega ^{ab}\wedge e^{c}\wedge
e^{d}\varepsilon _{abcd}+\int D\delta \omega ^{ab}\wedge B_{ab}+\frac{1}
{l_{p}}\int \widetilde{B}_{a}\wedge \delta \omega ^{ab}\wedge e_{b}=0 .
\end{equation}
Integral by parts, we have:
\begin{equation}
\frac{1}{\kappa}\int \delta \omega ^{ab}\wedge De^{c}\wedge
e^{d}\varepsilon _{abcd}+\int \delta \omega ^{ab}\wedge DB_{ab}  +\frac{1}{2 l_{p}}\int \delta \omega ^{ab}\wedge \left( \widetilde{B}%
_{a}\wedge e_{b}-\widetilde{B}_{b}\wedge e_{a}\right) =0 .
\end{equation}
\end{widetext}

From Eq. (\ref{RBrelation}), we further have $DB_{ab}=\frac{1}{\theta \pi}DR_{ab}=0$. Using the definition $De^{c}=T^{c}$, we finally obtain:

\begin{equation}
\frac{1}{\kappa}T^{c}\wedge e^{d}\varepsilon _{abcd}+\frac{1}{2l_{p}}
\left( \widetilde{B}_{a}\wedge e_{b}-\widetilde{B}_{b}\wedge e_{a}\right) =0.
\end{equation}
To derive the components of the above equation, we wedge $e^{f}$ on both sides:

\begin{eqnarray}
&&\frac{1}{4l_{p}}\left( \widetilde{B}_{aij}e^{i}\wedge e^{j}\wedge e_{b}-\widetilde{B}_{bij}e^{i}\wedge e^{j}\wedge e_{a}\right) \wedge e^{f} \nonumber \\
&=&-\frac{1}{2\kappa}T_{\ ij}^{c}e^{i}\wedge e^{j}\wedge e^{d}\wedge
e^{f}\varepsilon _{abcd},
\end{eqnarray}
Comparing with both sides, we obtain:
\begin{eqnarray}
&&\frac{1}{2l_{p}}\left( \widetilde{B}_{aij}\eta _{bd}\varepsilon ^{ijdf}-\widetilde{B}_{bij}\eta _{ad}\varepsilon ^{ijdf}\right)\nonumber\\
&=&-\frac{1}{\kappa}T_{\ ij}^{c}\delta _{abc}^{ijf}  \nonumber\\
&=&-\frac{2}{\kappa}\left( T_{\ ab}^{f}+T_{\ ca}^{c}\delta
_{b}^{f}-T_{\ cb}^{c}\delta _{a}^{f}\right)\equiv -\frac{2}{\kappa}Q_{\ ab}^{f},
\end{eqnarray}
where the coefficient 2 comes from $T_{\ ij}^{c}=-T_{\ ji}^{c}$. Note that with the Lorentz signature, contracting of $\varepsilon$ will lead to an extra minus sign, i.e., $\varepsilon ^{ijfd}\varepsilon _{abcd}=-\delta _{abc}^{ijf}$. According to the above definition, $Q_{\ ab}^{f} $ is also antisymmetric for the two lower indices. Now we lower down the index $f$ and multiply both sides by $\varepsilon ^{abcd}$, the above equation becomes:
\begin{equation}
\frac{1}{l_{p}}\varepsilon ^{abcd}\widetilde{B}_{a}^{\ ij}\varepsilon_{ijbf} =-\varepsilon ^{abcd}\frac{2}{\kappa}Q_{fab}.
\end{equation}
Expand the contraction of the total antisymmteric tensor $\varepsilon ^{abcd}\varepsilon_{ijbf}=\delta _{ijf}^{acd}$, we obtain:
\begin{equation}
\frac{1}{l_{p}}\left( \widetilde{B}_{f}^{\ cd}+\widetilde{B}_{a}^{%
\ ac}\delta _{f}^{d}-\widetilde{B}_{a}^{\ ad}\delta
_{f}^{c}\right) =-\varepsilon ^{abcd}\frac{1}{\kappa}Q_{fab}. \label{componentsofB} 
\end{equation}
By contracting $f$ and $c$, we further have:
\begin{equation}
-\frac{2}{l_{p}}\widetilde{B}_{a}^{\ ad}=-\varepsilon ^{abcd}\frac{1}{\kappa}Q_{abc} .
\end{equation}

Since $\varepsilon ^{abcd}\eta _{ac}T_{\ fb}^{f}=0$, we can simplify the above equation as:
\begin{equation}
\frac{1}{l_{p}}\widetilde{B}_{a}^{\ ad}=\varepsilon ^{abcd}\frac{1}{2\kappa}T_{abc} .
\end{equation}
Put this expression into Eq. (\ref{componentsofB}), we finally get the components of $\widetilde{B}_{f}$:
\begin{equation}
\frac{1}{l_{p}}\widetilde{B}_{f}^{\ cd}=-\frac{1}{\kappa}\varepsilon
^{abcd}Q_{fab}+\frac{1}{2\kappa}\left( \varepsilon ^{abgd}\delta
_{f}^{c}-\varepsilon ^{abgc}\delta _{f}^{d}\right) T_{abg}.
\end{equation}

Thus, we end up with:
\begin{eqnarray}
\frac{1}{l_{p}}\widetilde{B}_{a}^{\ ij}\varepsilon _{ijkl} &=&\frac{2}{%
\kappa}\delta _{kl}^{ij}Q_{aij}-\frac{1}{\kappa}\delta
_{akl}^{bcd}T_{bcd} \nonumber\\
&=&\frac{4}{\kappa}Q_{akl}-\frac{6}{\kappa}T_{\left[ akl\right] }.
\end{eqnarray}

Apply the relation between the torsion
and contorsion ($T^{a}=K^{a}_{\ b}\wedge e^{b}$, and the components $K_{abc}=-\frac{1}{2}(T_{abc}+T_{bca}-T_{cab})$. As seen in Appendix \ref{structure}), we can further simplify the above expression as:
\begin{eqnarray}
\frac{1}{l_{p}}\widetilde{B}_{a}^{\ ij}\varepsilon _{ijkl} &=&\frac{2}{\kappa}\left( T_{akl}-T_{kla}-T_{lak}\right) +\frac{4}{\kappa}\left(
T^{b}_{\ bk}\eta _{al}-T^{b}_{\ bl}\eta _{ak}\right)  \nonumber\\
&=&\frac{4}{\kappa}K_{kla}+\frac{4}{\kappa}\left( T^{b}_{\ bk}\eta
_{al}-T^{b}_{\ bl}\eta _{ak}\right). \label{BforewithoutM} 
\end{eqnarray}

Actually $\widetilde{B}_{f}$ can be written in a more compact form:
\begin{equation}
\widetilde{B}_{a}=\frac{l_{p}}{\kappa}\varepsilon_{abcd}e^{b}\wedge K^{cd}, \label{Bform}
\end{equation}
The full  details of the above derivation can be found in Appendix \ref{calofB}. 

(4)Variation with respect to the last variable $e$ gives:
\begin{eqnarray}
\frac{1}{\kappa}\int R^{ab}\wedge e^{c}\wedge \delta
e^{d}\varepsilon _{abcd}+\frac{1}{l_{p}}\int \widetilde{B}_{d}\wedge
D\delta e^{d} =0,
\end{eqnarray}
which leads to:
\begin{equation}
\frac{1}{\kappa}R^{ab}\wedge e^{c}\varepsilon _{abcd}-\frac{1}{l_{p}}D\widetilde{B}_{d}=0.
\end{equation}
Decompose $R$ as $R=\widetilde{R}+\widetilde{D}K+K\wedge K$. (See more details in Appendix \ref{structure}, where $\widetilde{D}$ is covariant derivative with Christofell connection $\Gamma$ and $\widetilde{R}$ is the torsion-free curvature tensor written in terms of $\Gamma$.) In such a way, we can decompose the above equation as:
\begin{widetext}
\begin{equation}
\frac{1}{\kappa}\left( \widetilde{R}^{ab}+K_{\ g}^{a}\wedge
K^{gb}\right) \wedge e^{c}\varepsilon _{abcd}+\frac{1}{l_{p}}K^{a}_{\ d}\wedge\widetilde{B}_{a} +\frac{1}{\kappa}\widetilde{D}K^{ab}\wedge e^{c}\varepsilon _{abcd}-\frac{1}{l_{p}}\widetilde{D}\widetilde{B}_{d}=0, \label{EwithoutM}
\end{equation}
where we also separate the covariant derivative of $\widetilde{B}_{d}$ into two parts $D\widetilde{B}_{d}=\widetilde{D}\widetilde{B}_{d}-K^{a}_{\ d}\wedge\widetilde{B}_{a}$. By substituting Eq. (\ref{Bform}), we can finally simplify the above equation as:
\begin{equation}
\frac{1}{\kappa}\left( \widetilde{R}^{ab}+K_{\ g}^{a}\wedge
K^{gb}\right) \wedge e^{c}\varepsilon _{abcd}+\frac{1}{\kappa}\varepsilon _{abmn}e^{b}\wedge K^{mn}\wedge K^{a}_{\ d}=0. \label{EwithoutM2}
\end{equation}
We note that the last two terms of Eq. (\ref{EwithoutM}) cancel out 
and $e^{c}$,$\varepsilon _{abcd}$ can be taken out from $\widetilde{D}$ since $\widetilde{D}e=\widetilde{D}\varepsilon _{abcd}=0$. To derive the components of the above equation, just wedge $e^{f}$ on both sides:
\begin{equation}
-\frac{1}{2\kappa}\widetilde{R}_{\ \ ij}^{ab}\delta _{abd}^{ijf}-\frac{1}{\kappa}K_{\ ci}^{a}K_{\ \ j}^{cb}\delta _{abd}^{ijf}+\frac{1}{\kappa}\delta _{amn}^{ijf}K^{mn}_{\ \ \ i}K_{\ dj}^{a}=0, \label{components}
\end{equation}
where we replace the index from $g$ to $c$ in the second term. The first term in left hand side is the usual Einstein tensor:
\begin{eqnarray}
\widetilde{R}_{\ \ ij}^{ab}\delta _{abd}^{ijf}&=&2\widetilde{R}_{\ \ ab}^{ab}\delta
_{d}^{f}+2\widetilde{R}_{\ \ bd}^{fb}+2\widetilde{R}_{\ \ da}^{af}=2\widetilde{R}\delta _{\ d}^{f}-4\widetilde{R}_{d}^{f}\equiv -4\widetilde{G}_{\ d}^{f}  .
\end{eqnarray}
The second term can be expressed as:
\begin{eqnarray}
K_{\ ci}^{a}K_{\ \ j}^{cb}\delta _{abd}^{ijf} &=& K_{\ %
ca}^{a}K_{\ \ b}^{cb}\delta _{d}^{f}+K_{\ cb}^{f}K_{\ \ %
d}^{cb}+K_{\ cd}^{a}K_{\ \ a}^{cf} -K_{\ cb}^{a}K_{\ \ a}^{cb}\delta _{d}^{f}-K_{\ %
ca}^{a}K_{\ \ d}^{cf}-K_{\ cd}^{f}K_{\ \ a}^{ca}  \nonumber
\\
&=&K_{\ ca}^{a}K_{\ \ b}^{cb}\delta _{d}^{f}+2K_{\ %
cb}^{f}K_{\ \ d}^{cb}-K_{\ cb}^{a}K_{\ \ a}^{cb}\delta
_{d}^{f}-2K_{\ ca}^{a}K_{\ \ d}^{cf}.\nonumber\\
\end{eqnarray}

Recall the definition of contorsion, we have $K_{\ ca}^{a}=-\frac{1}{2}\left( T_{\ ca}^{a}+T_{ca}^{\ \  a}-T_{a\ c}^{\ a}\right) =-T_{\ ca}^{a}$. Without causing confusion, we define $T_{\ ac}^{a}\equiv T_{c}$ here to simplify the above expression as:
\begin{equation}
K_{\ ci}^{a}K_{\ \ j}^{cb}\delta _{abd}^{ijf}=-\left(
T_{a}T^{a}+K_{abc}K^{bca}\right) \delta _{d}^{f}+2K_{\ cb}^{f}K_{\ 
\ d}^{cb}-2T_{c}K_{\ \ d}^{cf}.
\end{equation}

The last term can be expressed as:
\begin{equation}
\delta _{amn}^{ijf}K^{mn}_{\ \ \ i}K_{\ dj}^{a}=2\left(K^{jf}_{\ \ \ i}K_{\ dj}^{i}+K^{ij}_{\ \ \ i}K_{\ dj}^{f}+K^{fi}_{\ \ \ i}K_{\ dj}^{j}\right)=2(K^{fab}K_{dba}+T^{a}K^{f}_{\ da}-T^{f}T_{d}).
\end{equation}

By substituting all these expressions into Eq. (\ref{components}), we finally obtain:
\begin{eqnarray}
\widetilde{G}_{fd}&=&-\frac{1}{2}\left( T_{a}T^{a}+K_{abc}K^{bca}\right) \eta _{fd}+K_{fcb}K_{\ \ d}^{cb}-T_{c}K_{\ fd}^{c} -K_{fab}K_{d}^{\ ba}-T_{a}K_{fd}^{\ \ a}+T_{f}T_{d} \nonumber\\
&=&-\frac{l_{p}^2}{2}\left( \sigma_{a}\sigma^{a}+\Sigma_{abc}\Sigma^{bca}\right) \eta _{fd}+l_{p}^2\left(\Sigma_{fcb}\Sigma_{\ \ d}^{cb}-\sigma_{c}\Sigma_{\ fd}^{c} -\Sigma_{fab}\Sigma_{d}^{\ ba}-\sigma_{a}\Sigma_{fd}^{\ \ a}+\sigma_{f}\sigma_{d}\right),
 \label{totalGwithoutM}
\end{eqnarray}
where $\sigma _{a}=\sigma _{\ ba}^{b},$ and $\Sigma _{aij}=-\frac{1}{2}\left( \sigma _{aij}+\sigma _{ija}-\sigma _{jai}\right)$. Symmetrize the index $f$ and $d$, we can derive a generalized Einstein equation in the presence of loop-like extensive matter field:
\begin{eqnarray}
\frac{1}{\kappa}\widetilde{G}_{fd} = -\frac{1}{2\pi}\left( \sigma
_{a}\sigma ^{a}+\Sigma _{abc}\Sigma ^{bca}\right) \eta _{fd} +\frac{1}{2\pi}\left( \Sigma _{fab}\Sigma _{\ \ d}^{ab}+\Sigma
_{dab}\Sigma _{\ \ f}^{ab}\right) +\frac{1}{2\pi}\sigma _{a}\left( \sigma_{fd}^{\ \ a}+\sigma _{df}^{\ \ a}\right) -\frac{1}{\pi}\Sigma _{fab}\Sigma
_{d}^{\ ba}+\frac{1}{\pi}\sigma _{f}\sigma _{d}.\nonumber\\ \label{GwithoutM}
\end{eqnarray}

Surprisingly, if we interpret the right hand side of the above equation as the energy-momentum tensor of loop-like extensive matter, its trace vanishes, indicating the vanishing of scalar curvature in the above equation.
\begin{eqnarray}
\eta^{fd}\widetilde{G}_{fd}=-R&=&-2\left( T_{a}T^{a}+K_{abc}K^{bca}\right)
 +\left( K_{dab}K_{\ \ d}^{ab}\right) +T_{a}T^{a}-K_{fab}K^{fba}+T_{f}T^{f} \nonumber\\
&=&-2\left( T_{a}T^{a}+K_{abc}K^{bca}\right)+2T_{a}T^{a}+K_{abc}K^{bca}-K_{abc}K^{acb}=0.
\end{eqnarray}

Finally, the anti-symmetric part of Eq. (\ref{totalGwithoutM}) further requires:
\begin{equation}
\frac{1}{2}\left( \sigma _{fab}\sigma _{\
 \ d}^{ab}-\sigma _{dab}\sigma _{\ \ 
f}^{ab}\right) +\sigma _{\ ab}^{a}\left( \sigma _{fd }^{\ \ b}-\sigma _{df}^{\ \ b}\right) =0. \label{antiwithoutM2}
\end{equation}
\end{widetext}
Unlike the usual Einstein-Carton theory, it does not vanish automatically. 
However, it can be satisfied if we assume the internal loop current $\sigma _{abc}$ is totally antisymmetric. Of course, the most general solution of the above equation is still unclear and we will leave this problem in our future work. 

To this end, we see that the loop-like extensive matter has several intriguing features which make it to be a very promising candidate of dark matter. (a) First of all, there would be no other interactions between loop-like extensive matter and the normal matter except gravity. This is simply because loop-like extensive objects will not couple to the 1-form gauge fields and they will not participate all the interactions in Standard Model. (b) The energy-momentum tensor of loop-like extensive matter does not contribute to scalar curvature and such a unique property makes it very hard to be observed in a local way. Nevertheless, the existences of loop-like extensive matter will still greatly influence the large scale structure of galaxies in the universe. (c) Finally, the loop-like extensive objects are the most natural topological excitations in $3+1$D. Actually all point like objects should be regulated as loop-like objects at Planck scale, and this is one of the most important motivation in super-string theory. However, the loop-like extensive objects we proposed in this paper are very different from the critical string in super-string theory and they can even exist in $3+1$D without super symmetry.(This is because these topological excitations only rely on the topological sectors of the background space-time manifold, which is very different from the critical string that couples to the metric of the background space-time manifold.) Thus, in our framework, dark matter is more fundamental than the normal matter, and it is not a surprise why dark matter should dominate the universe. Below, we will discuss the origin of fermionic normal matter and derive the generalized Einstein equation in the presence Dirac field.

\section{Generalized Einstein equation with Dirac field}
\label{Dirac}
In this section, we will discuss the most general action including fermionic normal matter field, i.e., the Dirac field. Below we will consider the following total action: 
\begin{equation}
S'=S_{top}+S_{Dirac}-\frac{\theta \pi}{2}\int B^{ab}\wedge B_{ab}-\int \sigma
^{a}\wedge \widetilde{B}_{a} ,
\end{equation}
where $S_{top}$ defined in Eq. (\ref{S_top}), and $S_{Dirac}$ reads:
\begin{eqnarray}
S_{Dirac} &=&\frac{1}{3!}\int \varepsilon_{abcd} e^{a}\wedge e^{b}\wedge e^{c}\wedge \left(\frac{1}{2}\overline{\psi}\gamma
^{d} \nabla\psi+\left( h.c.\right)\right) \nonumber\\
&&+\frac{1}{4!}\int \varepsilon_{abcd} e^{a}\wedge e^{b}\wedge e^{c}\wedge e^{d}m \overline{\psi }\psi. \label{Sdirac2}
\end{eqnarray}
Here we use the following gamma matrix representation:
\begin{equation}
\gamma ^{0}=\left( 
\begin{array}{cc}
0 & -i \\ 
-i & 0%
\end{array}%
\right) ,\ \gamma ^{j}=\left( 
\begin{array}{cc}
0 & -i\sigma ^{j} \\ 
i\sigma ^{j} & 0%
\end{array}\right),
\end{equation}
and $\overline{\psi}$ is defined as $\overline{\psi}=i\psi^{\dagger} \gamma^{0}$. $\nabla$ is the usual covariant derivative for spinor field:
\begin{equation}
\nabla=d+\Omega=d+\Pi _{ab}\omega^{ab}.
\end{equation}
with $\Pi _{ab}=\frac{1}{8}\left[ \gamma _{a},\gamma _{b}\right] $.

Here we express the Dirac action in a topological invariant way, which indicates the topological origination of fermions. In a recent work, it was proposed that all elementary fermions can be regarded as topological Majorana zero modes at Planck scale. However, since topological Majorana zero modes must attach onto linked closed-loops in $3+1$D, it is very natural to assume all these topological zero modes arise from linked $e$-loops. After the condensation of $\omega$-loop, the linked $e$-loops will be confined at Planck scale and they can naturally be regarded as point-like particles. Of course, the action of bosonic matter fields, i.e.. gauge fields and Higgs field can not be written in such a topological invariant way, which suggests that the bosonic matter are not fundamental and they should be regarded as the collective motion of fermionic Dirac fields. How to derive the whole Standard Model from an underlying topological action is a very important question and we will leave this problem to our future work.

It is easy to verify that we can convert Eq. (\ref{Sdirac2}) to the conventional Dirac action by applying the following identities:
\begin{eqnarray}
\varepsilon_{abcd} e^{a}_{\mu}e^{b}_{\nu} e^{c}_{\rho}&=&\det(e)\varepsilon_{\mu\nu\rho\sigma} e^{\sigma}_{d}, \nonumber\\
\varepsilon_{abcd} e^{a}_{\mu}e^{b}_{\nu} e^{c}_{\rho}e^{d}_{\sigma}&=&\det(e)\varepsilon_{\mu\nu\rho\sigma},\label{identitye} 
\end{eqnarray}
which lead to:
\begin{eqnarray}
S_{Dirac}=-\int d^4x e\left[ \frac{1}{2}\overline{\psi }\gamma
^{a}e_{a}^{\mu }\left( \partial _{\mu }+\Omega _{\mu }\right) \psi +\left(
h.c.\right) + m\overline{\psi }\psi \right],
\nonumber\\
\end{eqnarray}
where $e\equiv \det(e^{a}_{\mu})$. 

Below we will perform the variational calculations for total action $S^\prime$. 

(1)Variation with respect to $B$ gives:
\begin{equation}
R^{ab}=\theta \pi B^{ab},\ \text{in terms of components}\  R_{\mu \nu }^{ab}=\theta \pi B_{\mu \nu
}^{ab}.
\end{equation}

(2)Variation with respect to $\widetilde{B}$ gives:
\begin{equation}
T^{a}=l_{p}\sigma ^{a},\ \text{in terms of components} \ \ T_{\mu \nu }^{a}=l_{p}\sigma
_{\mu \nu }^{a}.
\end{equation}

(3)Variation with respect to $\overline{\psi }$ gives:
\begin{eqnarray}
&&\frac{1}{3!}  \varepsilon_{abcd} e^{a}\wedge e^{b}\wedge e^{c}\wedge \gamma ^{d} \nabla\psi 
+\frac{1}{4!} \varepsilon_{abcd} e^{a}\wedge e^{b}\wedge e^{c}\wedge e^{d} m\psi \nonumber\\
&&-\frac{1}{4}\varepsilon_{abcd}T^{a}\wedge e^{b}\wedge e^{c} \gamma^{d}\psi =0.
\end{eqnarray}
The detailed derivation of above equation can be found in Appendix \ref{deriofphi}. To obtain the components of the above equation, we can apply the identity Eq. (\ref{identitye}) and finally have:
\begin{equation}
\gamma ^{a}e_{a}^{\mu }\left( \partial _{\mu }+\Omega _{\mu }\right) \psi
+m\psi -\frac{1}{2}T_{\ ab}^{a}\gamma ^{b}\psi =0.  \label{EOMofdirac}
\end{equation}
We note that the components of $-\frac{1}{4}\varepsilon_{abcd}T^{a}\wedge e^{b}\wedge e^{c}\gamma^{d}\psi$ can be expressed as:
\begin{equation}
-\frac{1}{8}\varepsilon_{abcd}\varepsilon^{\mu\nu\rho\sigma}T^{a}_{\mu\nu}e^{b}_{\rho}e^{c}_{\sigma}\gamma^{d}\psi
=\frac{1}{2}e e^{\mu}_{a}e^{\nu}_{d}T^{a}_{\mu\nu}\gamma^{d}\psi=\frac{1}{2}eT^{a}_{\ ab}\gamma^{b}\psi.
\end{equation}
The components of other terms can also be derived in a similarly way. We see that this EOM is very different from the usual Einstein-Cartan theory with Dirac field. Because in our case torsion is proportional to the loop source and has nothing to do with fermion current, there would be no fermion coupling term in Eq. (\ref{EOMofdirac}). 
If we separate the spin connection $\omega^{ab}$ into $\Gamma^{ab}$ and $K^{ab}$, the EOM of the Dirac field can further be simplified as:
\begin{eqnarray}
&&\gamma ^{a}e_{a}^{\mu }\left( \partial _{\mu }+\Pi^{cd}(\Gamma _{cd\mu }+K_{cd\mu })\right) \psi+m\psi -\frac{1}{2}T_{\ ab}^{a}\gamma ^{b}\psi \nonumber\\
&=&\gamma ^{a}e_{a}^{\mu }\left( \partial _{\mu }+\Pi^{cd}\Gamma _{cd\mu }\right) \psi+m\psi+\frac{1}{4}K_{acd}\gamma ^{acd}\psi\nonumber\\
&&+\frac{1}{2}K_{\ da}^{a}\gamma ^{d}\psi-\frac{1}{2}T_{\ ab}^{a}\gamma ^{b}\psi \nonumber\\
&=&\gamma ^{a}e_{a}^{\mu }\left( \partial _{\mu }+\Pi^{cd}\Gamma _{cd\mu }\right) \psi+m\psi -\frac{l_{p}}{8}\sigma_{abc}\gamma ^{abc}\psi=0
\end{eqnarray}
where we use the identity:
\begin{eqnarray}
\gamma^{a}\Pi^{cd}&=&\frac{1}{2}\left(\gamma^{a}\Pi^{cd}+\Pi^{cd}\gamma^{a}\right)+\frac{1}{4}(\eta^{ac}\gamma^{d}-\eta^{ad}\gamma^{c}) \nonumber\\
&=&\frac{1}{4}\gamma^{acd}+\frac{1}{4}(\eta^{ac}\gamma^{d}-\eta^{ad}\gamma^{c}),
\end{eqnarray}
with $\gamma ^{abc}=\gamma ^{\lbrack a}\gamma^{b}\gamma ^{c]}$ and $K_{\ ba}^{a}=T_{\ ab}^{a}$.

Interestingly, in the presence of dark matter(loop current) background, the extra term $-\frac{l_{p}}{8}\sigma_{abc}\gamma ^{abc}$ can be regarded as a small correction to the fermion mass. In general, the loop current background could have a non-uniform distribution, which implies a small fluctuation of fermion mass with dark matter background. Of course, as the correction is proportional to the Planck length $l_p$, it is not clear if such a tiny change of fermion mass can be measured by current experiment or not. 



Next we can further compute the variation of $S'$ with respect to $e$ and $\omega$. We first consider the variation of Dirac field with respect to $e$ and compute energy momentum tensor $t_{d}$ for Dirac field.  
\begin{widetext} 
\begin{eqnarray}
\delta _{e}S_{Dirac}&=&\frac{1}{2!}\int \varepsilon_{abcd} \delta e^{a}\wedge e^{b}\wedge e^{c}\wedge \left(\frac{1}{2}\overline{\psi}\gamma^{d} \nabla\psi+\left( h.c.\right)\right) +\frac{1}{3!}\int \varepsilon_{abcd}\delta e^{a}\wedge e^{b}\wedge e^{c}\wedge e^{d}m \overline{\psi }\psi \nonumber\\
&=&\frac{1}{2!}\int \varepsilon_{abcd} \left(\frac{1}{2}\overline{\psi}\gamma^{a} \nabla\psi+\left( h.c.\right)\right) \wedge e^{b}\wedge e^{c}\wedge \delta e^{d}+\frac{1}{3!}\int \varepsilon_{abcd} e^{a}\wedge e^{b}\wedge e^{c}\wedge \delta e^{d}m \overline{\psi }\psi  \nonumber\\
&\equiv & -\int  t_{d}\wedge\delta e^{d},
\end{eqnarray}
where
\begin{equation}
t_{d}=-\frac{1}{2!} \varepsilon_{abcd} \left(\frac{1}{2}\overline{\psi}\gamma^{a} \nabla\psi+\left( h.c.\right)\right) \wedge e^{b}\wedge e^{c}
-\frac{1}{3!}\varepsilon_{abcd} e^{a}\wedge e^{b}\wedge e^{c} m \overline{\psi }\psi. \label{EMtensor1}
\end{equation}
To simplify the energy momentum tensor, we can make use of the EOM for Dirac field. We wedge $e^{f}$ on the right to make it a 4-form and derive the components of the above equation as:
\begin{eqnarray}
&&-\frac{1}{2!} \varepsilon_{abcd}\varepsilon^{ibcf} \left(\frac{1}{2}\overline{\psi}\gamma^{a} \nabla_{i}\psi+\left( h.c.\right)\right)
-\frac{1}{3!}\varepsilon_{abcd}\varepsilon^{abcf} m \overline{\psi }\psi\nonumber\\
&=&\left(\frac{1}{2}\overline{\psi}\gamma^{a} \nabla_{a}\psi+\left( h.c.\right)+m \overline{\psi }\psi\right)\delta^{f}_{d}-\left(\frac{1}{2}\overline{\psi}\gamma^{f} \nabla_{d}\psi+\left( h.c.\right)\right)\nonumber\\
&=&-\left(\frac{1}{2}\overline{\psi}\gamma^{f} \nabla_{d}\psi+\left( h.c.\right)\right). \label{twedgeef}
\end{eqnarray}
The first term of the second line vanishes on shell. 
On the other hand, we find the component of $-\frac{1}{3!}\varepsilon _{gabc}\left( \frac{1}{2}\overline{\psi }
\gamma ^{g}\nabla_{d}\psi
+\left( h.c.\right) \right) e^{a}\wedge e^{b}\wedge e^{c}$ can also be written as:
\begin{equation}
-\frac{1}{3!}\varepsilon _{gabc}\varepsilon^{abcf}\left( \frac{1}{2}\overline{\psi }\gamma ^{g}\nabla_{d}\psi
+\left( h.c.\right) \right) =-\left(\frac{1}{2}\overline{\psi}\gamma^{f} \nabla_{d}\psi+\left( h.c.\right)\right),
\end{equation}
which is exactly the same as Eq. (\ref{twedgeef}). Thus we can simplify the energy momentum tensor as:
\begin{eqnarray}
t_{d} &=&-\frac{1}{3!}\varepsilon _{fabc}\left( \frac{1}{2}\overline{\psi }
\gamma ^{f}\nabla_{d}\psi
+\left( h.c.\right) \right) e^{a}\wedge e^{b}\wedge e^{c} \equiv \frac{1}{3!}\varepsilon _{fabc}t_{\ d}^{f}e^{a}\wedge e^{b}\wedge e^{c} ,
\end{eqnarray}
where $t_{\ d}^{f}=-\frac{1}{2}\overline{\psi }\gamma ^{f}e_{d}^{\mu }\left(
\partial _{\mu }+\Omega _{\mu }\right) \psi +\left( h.c.\right) .$

Similarly, we can also consider the variation of Dirac field with respect to $\omega$: 
\begin{eqnarray}
\delta _{\omega }S_{Dirac} &=&\frac{1}{3!}\frac{1}{2}\int \varepsilon _{cdfg} e^{c}\wedge e^{d}\wedge e^{f} \wedge\delta\omega ^{ab}\overline{\psi }\left( \gamma ^{g}\Pi
_{ab}+\Pi _{ab}\gamma ^{g}\right) \psi \\
&=&\frac{1}{3!}\frac{1}{2}\int \varepsilon _{gcdf}\delta \omega ^{ab}\wedge e^{c}\wedge e^{d}\wedge e^{f}\overline{\psi }\left(\gamma ^{g}\Pi _{ab}+\Pi _{ab}\gamma ^{g}\right) \psi \nonumber\\
& \equiv & -\int \delta \omega
^{ab}\wedge \widetilde{T}_{ab},
\end{eqnarray}
where
\begin{equation}
\widetilde{T}_{ab}=-\frac{1}{3!}\frac{1}{2}\varepsilon _{gcdf}\overline{\psi }\left(
\gamma ^{g}\Pi _{ab}+\Pi _{ab}\gamma ^{g}\right) \psi e^{c}\wedge
e^{d}\wedge e^{f}=\frac{1}{3!}\widetilde{T}_{\ ab}^{g}\varepsilon
_{gcdf}e^{c}\wedge e^{d}\wedge e^{f}, 
\end{equation}
and $\widetilde{T}_{\ ab}^{g}=-\frac{1}{2}\overline{\psi }\left(
\gamma ^{g}\Pi _{ab}+\Pi _{ab}\gamma ^{g}\right) \psi=-\frac{1}{4}\overline{\psi }\gamma ^{\lbrack a}\gamma^{b}\gamma ^{c]}\psi$, which is totally antisymmetric with respect to all indices. 

Now we are ready to compute the variation with respect to $\omega$ and $e$ for the total action $S'$.

(4)Variation of the total action $S^\prime$ with respect to $\omega $ yields:
\begin{equation}
\delta _{\omega}S^\prime=\frac{1}{2\kappa}\int D\delta \omega ^{ab}\wedge e^{c}\wedge
e^{d}\varepsilon _{abcd}+\int D\delta \omega ^{ab}\wedge B_{ab}+\frac{1}{l_{p}}\int \widetilde{B}_{a}\wedge \delta \omega ^{ab}\wedge
e_{b}-\int \delta \omega ^{ab}\wedge \widetilde{T}_{ab}=0 .
\end{equation}
Similar as the case without Dirac filed, integral by parts leads to:
\begin{equation}
\frac{1}{\kappa}T^{c}\wedge e^{d}\varepsilon _{abcd}+\frac{1}{2l_{p}}\left( \widetilde{B}_{a}\wedge e_{b}-\widetilde{B}_{b}\wedge e_{a}\right)
=\widetilde{T}_{ab}. \label{BwithM}
\end{equation}
The solution of $\widetilde{B}_{a}$ to the above equation is:
\begin{equation}
\widetilde{B}_{d}=\frac{l_{p}}{\kappa}
\varepsilon _{abcd} e^{b} \wedge K^{cd}+
\frac{l_{p}}{2}\varepsilon _{abcd} e^{b}\wedge e^{f}\widetilde{T}_{f}^{\ cd} ,
\end{equation}
See in Appendix \ref{calofB} for full details.

(5)Finally we consider the variation of $S'$ with respect to the last variable $e$:

\begin{equation}
\delta _{e}S^\prime =\frac{1}{\kappa}\int R^{ab}\wedge e^{c}\wedge \delta
e^{d}\varepsilon _{abcd}+\frac{1}{l_{p}}\int \widetilde{B}_{d}\wedge
D\delta e^{d}-\int t_{d}\wedge \delta e^{d}=0,
\end{equation}
which gives:
\begin{equation}
\frac{1}{\kappa}R^{ab}\wedge e^{c}\varepsilon _{abcd}-\frac{%
1}{l_{p}}D\widetilde{B}_{d}=t_{d}.
\end{equation}
Again, decompose $R$ and the covariant derivative in to torsion-free part and torsion part, we obtain:
\begin{equation}
\frac{1}{\kappa}\left( \widetilde{R}^{ab}+K_{\ g}^{a}\wedge
K^{gb}\right) \wedge e^{c}\varepsilon _{abcd}+\frac{1}{l_{p}}\widetilde{B}_{a}\wedge K_{\ d}^{a}-t_{d} +\frac{1}{\kappa}\widetilde{D}K^{ab}\wedge e^{c}\varepsilon _{abcd}-\frac{1}{l_{p}}\widetilde{D}\widetilde{B}_{d}=0. \label{EwithM}
\end{equation}
By substituting the solution of $\widetilde{B}_{a}$, we can simplify the above equation as:
\begin{equation}
\frac{1}{\kappa}\left( \widetilde{R}^{ab}+K_{\ g}^{a}\wedge
K^{gb}\right) \wedge e^{c}\varepsilon _{abcd}+\frac{1}{\kappa}\varepsilon _{abmn}e^{b}\wedge K^{mn}\wedge K_{\ d}^{a}=t_{d}-\frac{1}{2}\varepsilon _{abmn}e^{b}\wedge e^{g}\wedge K_{\ d}^{a}\widetilde{T}_{g}^{\ mn}+\frac{1}{2}\varepsilon _{dbmn}e^{b}\wedge e^{g}\widetilde{D}\widetilde{T}_{g}^{\ mn}. \label{EwithM2}
\end{equation}
The components of the left hand side are the same as the case without Dirac field. The components of the right hand side can be obtained by wedge $e^{f}$ on the right side:
\begin{eqnarray}
&&\frac{1}{3!}\varepsilon _{gabc}\varepsilon ^{abcf}t^{g}_{\ d}-\frac{1}{2}\varepsilon _{abmn}\varepsilon^{bgif} K_{\ di}^{a}\widetilde{T}_{g}^{\ mn} +\frac{1}{2}\varepsilon _{dbmn}\varepsilon^{bgif} \widetilde{D}_{i}\widetilde{T}_{g}^{\ mn} \nonumber\\
&=& t^{f}_{\ d}-\left(K_{\ di}^{a}\widetilde{T}_{a}^{\ if}+K_{\ da}^{a}\widetilde{T}_{g}^{\ fg}+K_{\ di}^{f}\widetilde{T}_{g}^{\ gi}\right)+\left(\widetilde{D}_{i}\widetilde{T}_{d}^{\ if}+\widetilde{D}_{i}\widetilde{T}_{g}^{\ gi}\delta^{f}_{d}+\widetilde{D}_{d}\widetilde{T}_{g}^{\ fg}\right)\nonumber\\
&=& t^{f}_{\ d}+\widetilde{T}^{abf}K_{dab}+\widetilde{D}_{i}\widetilde{T}_{\ \ d}^{if},
\end{eqnarray}
where we use the total antisymmetric property of $\widetilde{T}$. 

Thus, the components of Eq. (\ref{EwithM2}) read:
\begin{eqnarray}
\widetilde{G}_{fd} &=&-\frac{1}{2}\left( T_{a}T^{a}+K_{abc}K^{bca}\right) \eta _{fd}+K_{fcb}K_{\ \ d}^{cb}-T_{c}K_{\ fd}^{c} -K_{fab}K_{d}^{\ ba}-T_{a}K_{fd}^{\ \ a}+T_{f}T_{d}+\frac{\kappa}{2} \left( t_{fd}+\widetilde{T}_{abf}K_{d}^{\ ab}+\widetilde{D}_{i}
\widetilde{T}_{\ fd}^{i}\right) \nonumber\\
&=&-\frac{l_{p}^2}{2}\left( \sigma_{a}\sigma^{a}+\Sigma_{abc}\Sigma^{bca}\right) \eta _{fd}+l_{p}^2(\Sigma_{fcb}\Sigma_{\ \ d}^{cb}-\sigma_{c}\Sigma_{\ fd}^{c} -\Sigma_{fab}\Sigma_{d}^{\ ba}-\sigma_{a}\Sigma_{fd}^{\ \ a}+\sigma_{f}\sigma_{d})\nonumber\\
&&+\frac{\kappa l_{p}}{2} \left( \widetilde{T}_{abf}\Sigma_{d}^{\ ab}+\widetilde{T}_{abf}\Sigma_{\ \ d}^{ab}\right)+\frac{\kappa}{2}\left(-\frac{1}{2}\overline{\psi }\gamma _{f}e_{d}^{\mu }\left( \partial _{\mu
}+\Gamma_{ab\mu }\Pi ^{ab}\right) \psi +\left( h.c.\right)+\widetilde{D}_{i}\widetilde{T}_{\ fd}^{i}\right), \label{EOMDirac}
\end{eqnarray}
where in the second step, we separate $t_{fd}$ into torsion free part and torsion part:
\begin{eqnarray}
t_{fd} &=&-\frac{1}{2}\overline{\psi }\gamma _{f}e_{d}^{\mu }\left( \partial _{\mu}+\Gamma_{ab\mu }\Pi ^{ab}\right) \psi -\frac{1}{2}K_{\ \ d}^{ab}\overline{\psi }\gamma _{f}\Pi _{ab}\psi-\frac{1}{2}\left( -\partial _{\mu}\overline{\psi }+\overline{\psi }\Gamma_{ab\mu }\Pi ^{ab}\right)\gamma _{f}e_{d}^{\mu } \psi-\frac{1}{2}K_{\ \ d}^{ab}\overline{\psi }\Pi _{ab}\gamma _{f}\psi \nonumber\\
&=&\left(-\frac{1}{2}\overline{\psi }\gamma _{f}e_{d}^{\mu }\left( \partial _{\mu
}+\Gamma_{ab\mu }\Pi ^{ab}\right) \psi +\left( h.c.\right)\right)+
\widetilde{T}_{abf}K_{\ \ d}^{ab}.
\end{eqnarray}

Finally, after symmetrizing the index $f$ and $d$, we end up with the generalized Einstein equation with both dark matter and Dirac field:

\begin{eqnarray}
\frac{1}{\kappa}\widetilde{G}_{fd} &=&-\frac{1}{2\pi}\left( \sigma
_{a}\sigma ^{a}+\Sigma _{abc}\Sigma ^{bca}\right) \eta _{fd}+\frac{1}{2\pi}\left( \Sigma _{fab}\Sigma _{\ \ d}^{ab}+\Sigma
_{dab}\Sigma _{\ \ f}^{ab}\right) +\frac{1}{2\pi}\sigma _{a}\left( \sigma
_{fd}^{\ \ a}+\sigma _{df}^{\ \ a}\right) -\frac{1}{\pi}\Sigma _{fab}\Sigma
_{d}^{\ ba}+\frac{1}{\pi}\sigma _{f}\sigma _{d}  \nonumber \\
&&+\frac{l_{p}}{4}\left[ \widetilde{T}_{abf}(\Sigma _{d}^{\ ab}+\Sigma _{\ \ d}^{ab})+\widetilde{T}_{abd}\left( \Sigma _{f}^{\ ab}+\Sigma
_{\ \ f}^{ab}\right) \right]-\frac{1}{8}\left[ \overline{\psi }\left( \gamma _{f}e_{d}^{\mu }+\gamma
_{d}e_{f}^{\mu }\right) \left( \partial _{\mu }+\Gamma_{ab\mu
}\Sigma ^{ab}\right) \psi +(h.c.)\right]. \label{GwithM}
\end{eqnarray}

The first line is the same as Eq. (\ref{GwithoutM}), which is the energy momentum tensor of the dark matter. The first term in the second line can be regarded as the coupling between the dark matter and Dirac field, and the second term in the second line is the usual energy momentum tensor of Dirac field. 

In addition, the antisymmetric part of Eq. (\ref{EOMDirac}) can be simplified as:
\begin{equation}
\frac{1}{2}\left( \sigma _{fcb}\sigma _{\ \ d}^{cb}-\sigma
_{dcb}\sigma_{\ \ f}^{cb}\right) +\sigma _{a}\left( \sigma _{fd}^{
\ \ a}-\sigma _{df}^{\ \ a}\right) =\frac{\pi }{4 }l_{p}(\sigma _{f}^{\ ab}\widetilde{T}_{dab}-\sigma _{d}^{\ 
ab}\widetilde{T}_{fab}). \label{antiwithM}
\end{equation}

The detailed derivation can be found in Appendix \ref{antisymmetric}. Again, this equation doesn't have a obvious solution. Even with totally antisymmetric internal loop current $\sigma_{abc}$, this equation are not satisfied automatically due to the additional terms on the right-hand side. However, since the right-hand side is proportional to $l_p$, the totally antisymmetric internal loop current $\sigma_{abc}$ is still an approximate solution in the limit $l_p \rightarrow 0$.


\section{Other possible topological terms and the uniqueness of generalized Einstein equation}
\label{general} 
In the loop condensation phase, due to the breaking of 2-form and 1-form gauge transformations, we can in principle consider more general topological terms which are invariant under local Lorentz transformation. It is very important to understand the effect of these terms to justify the uniqueness of the generalized Einstein equation. Below we will consider the following most general topological invariant action(in the absence of Dirac field): 

\begin{eqnarray}
S &=&\frac{1}{2\kappa}\int \varepsilon _{abcd}R^{ab}\wedge e^{c}\wedge e^{d}+\int
R^{ab}\wedge B_{ab}+\frac{1}{l_{p}}\int T^{a}\wedge \widetilde{B}_{a}-\frac{\theta \pi }{2}\int B^{ab}\wedge B_{ab}-\int \sigma
^{a}\wedge \widetilde{B}_{a} \nonumber \\
&&-\frac{\lambda }{2}\int
T^{a}\wedge T_{a}+\frac{\psi }{2}\int R^{ab}\wedge e_{a}\wedge e_{b}.\label{generalS} 
\end{eqnarray}
\end{widetext}
The above action includes all possible topological invariant terms except $\widetilde{B}^{a}\wedge \widetilde{B}
_{a}$. This is because the loops coupled with $B$ and $\widetilde{B}$ can not be condensed simultaneously due to the presence of $Ree$ term, as discussed in section \ref{loopsource}. 
Similar as before, variation with respect to $B$ and $\widetilde{B}$ gives:
\begin{equation}
R^{ab}=\theta\pi B^{ab}, \ \ \ T^{a}=l_{p}\sigma ^{a}.
\end{equation}

Variation with  respect to $\omega$ gives:
\begin{equation}
\frac{1}{\kappa} T^{c}\wedge e^{d}\varepsilon _{abcd}+(\frac{1}{l_{p}}\widetilde{B}+(\psi -\lambda
)T)_{[a}\wedge e_{b]}=0 .
\end{equation}
As seen in Appendix \ref{calofB}, the general solution of $\widetilde{B}_{a}$ reads:
\begin{equation}
\widetilde{B}_{a}=-l_{p}(\psi -\lambda )T_{a}+\frac{l_{p}}{\kappa}\varepsilon _{abcd}e^{b}\wedge K^{cd} ,\label{generalB}
\end{equation}

Finally, variation with respect to $e^{d}$ gives:
\begin{widetext}
\begin{eqnarray}
&&\frac{1}{\kappa}\varepsilon _{abcd}R^{ab}\wedge e^{c}-\frac{1}{l_{p}}D\widetilde{B}_{d}+\lambda DT_{d}+\psi R_{ad}\wedge e^{a}\nonumber\\
&=&\frac{1}{\kappa}\varepsilon _{abcd} \left(\widetilde{R}^{ab}+\widetilde{D}K^{ab}+K_{\ g}^{a}\wedge K^{gb}\right)\wedge e^{c}-\frac{1}{l_{p}}\left(\widetilde{D}\widetilde{B}_{d}-\widetilde{B}_{a}\wedge K^{a}_{\ d}\right)+(\psi -\lambda )R_{ad}\wedge e^{a}, 
\end{eqnarray}
where in the second step we expand $R$ and the covariant derivative. We also apply the structure equation $DT_{d}=R^{a}_{\ b}\wedge e^{b}$ (See Appendix \ref{structure}). By substituting Eq. (\ref{generalB}) in, we finally have:
\begin{equation}
\varepsilon _{abcd} \widetilde{R}^{ab} \wedge e^{c}+\varepsilon _{abcd}K_{\ f}^{a}\wedge K^{fb}\wedge e^{c} +\varepsilon _{abmn}e^{b}\wedge K^{mn} \wedge K_{\ d}^{a}=0,
\end{equation}
\end{widetext}
which is exactly the same as Eq. (\ref{EwithoutM2}). By adding the Dirac field into the action Eq. (\ref{generalS}), we will still get the same field equation as Eq. (\ref{EwithM2}).

Thus we conclude that the generalized Einstein equation we derive here is very unique and additional topological invariant term will not  
change the EOM. 
Of course, the cosmology constant term can always be include in our framework naturally since it can also be rewritten as a topological invariant form with
$S_\Lambda=\frac{\Lambda}{4!}\int \epsilon_{abcd} e^a \wedge e^b \wedge e^c \wedge e^d $. 
Moreover, all the topological invariant terms are dimension 4 operators and the whole theory is still power counting renormalizable. All these nice properties suggest that topological invariant principle is more profound than general covariance principle.

\section{Conclusion and discussions}
In this paper, we begin with a 3+1D topological gravity theory which is super renormalizable, then we argue that Einstein gravity will naturally emerge from such a topological gravity theory via $\omega$-loop condensation. Although a rigorous way of treating loop condensation requires a new framework of string field theory\footnote{As loop like excitations arise from the underlying 3+1D TQFT can be regarded as a non-smooth string, therefore, we believe such a string field theory is very different from the usual super string theory which can only be defined in 10 or 11 dimensional spacetime} to describe loop like extensive excitations, the semiclassicial limit can still be achieved by simply adding a topological mass term for the 2-form gauge field $B$ which couples to $\omega$-loop. Remarkably, such a semi-classical theory is still super renormalizable. 

Moreover, our approach indicates that loop-like extensive objects could be the most general form of matter source(including both normal matter and dark matter) in 3+1D space time. In particular, those uncondensed (deconfined) loop-like excitations can be naturally interpreted as dark matter. By introducing a loop source term of (unlinked) $e$-loop(that is, the topological invariant coupling between $e$-loop current and the corresponding 2-form gauge field $\tilde{B}$), we obtain a generalized Einstein equation which naturally includes the dark matter sector. Surprisingly, we find that the trace of the energy-momentum tensor vanishes for dark matter. As a result, such kind of "conformal" dark matter will not contribute to scalar curvature.  On the other hand, we predict that the current of such kind of dark matter will naturally contribute to torsion.

In addition, we also investigate the essential difference between our theory and the usual Einstein-Carton theory by introducing Dirac fields. Interestingly, we find that the fermion current \textit{will not} contribute to torsion. From the renormalization group perspective, the traditional Einstein-Carton theory is inconsistent with the relativistic quantum field theory. This is because the torsion produced by Dirac spinor will lead to four fermion interactions, however, relativistic quantum field theory tells us that four fermion interactions are irrelevant in $3+1$D and they should not appear in the semiclassical limit(for any fixed background metric).   
Thus, we claim that all normal matters actually will not produce torsion while only dark matter can be the possible source of torsion. 
Such a strong prediction can be tested in future experiments. Finally, we consider all the allowed topological invaraint terms and we find all these terms will not change the generalized Einstein equation.

On the other hand, although all normal matter looks like point like particle, it has been argued for decades that there should be no point-like particle at Planck scale due to the UV divergence in quantum field theory. In usual super string theory, it was conjectured that all elementary particles are actually different vibration modes of critical strings at Planck scale. However, super string theory requires both super symmetry and extra space-time dimensions, but none of them are observed experimentally so far. Very recently, a topological scenario suggests that all elementary particles are actually topological zero modes of linked loop-like excitations. Such an approach naturally explains the origin of three generations of elementary particles and gives rise to the correct neutrino mass mixing angles. In our theory, when $\omega$-loop condenses, the linked $e$-loops will be confined and they can be regarded as fermionic particles if topological Majroana modes can be further attached on to these linked $e$-loops. 
Apparently, understanding the dynamics and quantum effect of loop-like extensive matter will become very important but challenging direction in the future. In particular, the notation of loop condensation might unify all fundamental forces and tell us the origin of four dimensional space-time! 

\section{Acknowledgement}
We would like to thank Yongshi Wu and Ming-Chung Chu for helpful discussions.
This work is supported by a grant from the Research Grants Council of the Hong Kong Special Administrative Region, China (Project No. AoE/P-404/18).

\appendix

\section{Structure equations}\label{structure}

Below we will list several useful identities and related definitions. 

For $R_{\ b}^{a}\in \Lambda_{\left( 1,1\right) }^{2}$, it is straightforward to verify that:
\begin{eqnarray}
DR &=&D\left( d\omega +\omega \wedge \omega \right) =d\left( d\omega +\omega
\wedge \omega \right) +\omega \wedge R-R\wedge \omega  \nonumber\\
&=&d\omega \wedge \omega -\omega \wedge d\omega +\omega \wedge d\omega
+\omega \wedge \omega -d\omega \wedge \omega -\omega \wedge \omega  
\nonumber\\
&=&0.
\end{eqnarray}

By replacing $\omega$ with the torsion free connection $\Gamma$ in the above equation, we have:
\begin{equation}
\widetilde{D}\widetilde{R}=0 ,
\end{equation}
where $\widetilde{D}$ is the covariant derivative with respect to Christofell connection $\Gamma$, and $\widetilde{R}$ is written in terms of $\Gamma$ , and the explict form of $\Gamma$ reads: 
\begin{eqnarray}
\Gamma_{\mu \nu }^{\rho }&=&\frac{1}{2}g^{\rho \lambda }\left(
\partial _{\mu }g_{\nu \lambda }+\partial _{\nu }g_{\lambda \mu }-\partial
_{\lambda }g_{\mu \nu }\right)\nonumber\\
&=&\frac{1}{2}e^{a\rho}e^{\lambda}_{b}(\partial _{\mu }e^{a}_{\nu}e_{a\lambda}+e^{a}_{\nu}\partial _{\mu }e_{a\lambda}+\partial _{\nu }e^{a}_{\mu}e_{a\lambda}\nonumber\\
&&+e^{a}_{\mu}\partial _{\nu }e_{a\lambda}-\partial _{\lambda }e^{a}_{\mu}e_{a\nu}-\partial _{\lambda }e^{a}_{\nu}e_{a\mu}).
\end{eqnarray}
Through a coordinate transformation, we can also write it with internal index:
\begin{eqnarray}
\Gamma_{\ b \mu}^{a}&=& e^{a}_{\rho} \partial_{\mu} e^{\rho}_{b}+e^{a}_{\rho} \Gamma_{\nu \mu }^{\rho }e^{\nu}_{b}\nonumber\\
&=& e^{a}_{\rho} \partial_{\mu} e^{\rho}_{b}+\frac{1}{2}e^{a\lambda}e^{\nu}_{b}(\partial _{\mu }e^{c}_{\nu}e_{c\lambda}+e^{c}_{\nu}\partial _{\mu }e_{c\lambda}+\partial _{\nu }e^{c}_{\mu}e_{c\lambda}\nonumber\\
&&+e^{c}_{\mu}\partial _{\nu }e_{c\lambda}-\partial _{\lambda }e^{c}_{\mu}e_{c\nu}-\partial _{\lambda }e^{c}_{\nu}e_{c\mu})\nonumber\\
&=&\frac{1}{2}[e^{\rho}_{b}(\partial _{\rho }e_{\mu}^{a}-\partial _{\mu }e_{\rho}^{a})+e^{a\rho}(\partial _{\mu }e_{b\rho}-\partial _{\rho}e_{b\mu})\nonumber\\
&&+e^{a\lambda}e^{\nu}_{b}e^{c}_{\mu}(\partial _{\nu }e_{c\lambda}-\partial _{\lambda }e_{c\nu})].
\end{eqnarray} 

For $T^{a}\in \Lambda _{\left( 1,0\right) }^{2}$, we have:
\begin{eqnarray}
DT^{a} &=&d\left( de^{a}+\omega _{\ b}^{a}\wedge e^{b}\right) +\omega
_{\ b}^{a}\wedge \left( de^{b}+\omega _{\ c}^{b}\wedge
e^{c}\right)  \\
&=&d\omega _{\ b}^{a}\wedge e^{b}-\omega _{\ b}^{a}\wedge
de^{b}+\omega _{\ b}^{a}\wedge de^{b}+\omega _{\ b}^{a}\wedge
\omega _{\ c}^{b}\wedge e^{c}  \nonumber \\
&=&R_{\ b}^{a}\wedge e^{b}.  \nonumber
\end{eqnarray}

Again, if we take $\widetilde{D}$ on $e^{a}$:
\begin{eqnarray}
\widetilde{D}e^{a} &=&de^{a}+\Gamma_{\ b}^{a}\wedge e^{b}
\\&=&\frac{1}{2}\left( \partial _{\mu }e_{\nu }^{a}-\partial _{\nu }e_{\mu }^{a}+%
\Gamma_{\ b\mu }^{a}e_{\nu }^{b}-\Gamma_{%
\ b\nu }^{a}e_{\mu }^{b}\right) dx^{\mu }\wedge dx^{\nu } \nonumber\\
&=&\frac{1}{2}e_{\rho }^{a}e_{c}^{\rho }\left( \partial _{\mu }e_{\nu
}^{c}-\partial _{\nu }e_{\mu }^{c}+\Gamma_{\ b\mu
}^{c}e_{\nu }^{b}-\Gamma_{\ b\nu }^{c}e_{\mu }^{b}\right)
dx^{\mu }\wedge dx^{\nu }  \nonumber \\
&=&\frac{1}{2}e_{\rho }^{a}\left( \Gamma_{\nu \mu }^{\rho }-%
\Gamma_{\mu \nu }^{\rho }\right) dx^{\mu }\wedge dx^{\nu }=0 .
\nonumber \label{freetorsion}
\end{eqnarray}

Thus, it is convenient to separate the spin connection into two pieces:

\begin{equation}
\omega _{\ b}^{a}=\Gamma_{\ b}^{a}+K_{\ b}^{a}, \label{separation}
\end{equation}
where $K_{\ b}^{a}$ is called contorsion. We can get the relation between $T$ and $K$ from Eq. (\ref{freetorsion}), which is:
\begin{equation}
T^{a}=De^{a}=\widetilde{D}e^{a}+K_{\ b}^{a}\wedge e^{b}=K_{\ b}^{a}\wedge e^{b},
\end{equation}
and the corresponding component form reads:
\begin{equation}
K_{ab \mu}=-\frac{1}{2}(e^{\rho}_{b}T_{a\rho\mu}+e^{\rho}_{a}T_{b\mu\rho}-e^{c}_{\mu}e^{\rho}_{a}e^{\sigma}_{b}T_{c\rho\sigma}).
\end{equation}

By using the decomposition (\ref{separation}), we can rewrite the curvature tensor into:
\begin{eqnarray}
R &=&d( \Gamma+K) +\left( \Gamma%
+K\right) \wedge \left( \Gamma+K\right) 
\nonumber
\\ \nonumber
&=&d\Gamma+dK+\Gamma\wedge \Gamma+%
\Gamma\wedge K+K\wedge \Gamma+K\wedge K  
\\ 
&=&\widetilde{R}+\widetilde{D}K+K\wedge K .
\end{eqnarray}

Finally, it is easy to show that $\varepsilon _{abcd}$ is a covariant constant:
\begin{eqnarray}
D\varepsilon _{abcd} &=&d\varepsilon _{abcd}-\omega _{\ a}^{f}\varepsilon_{fbcd}-\omega _{\ b}^{f}\varepsilon _{afcd}-\omega _{\ c}^{f}\varepsilon _{abfd}-\omega _{\ d}^{f}\varepsilon _{abcf} \nonumber\\
&=&0.
\end{eqnarray}

The first term $d\varepsilon _{abcd}$ is zero. For the left four terms, we can consider their components. Without loss of generality, we set $a=1,b=1,c=2,d=3$ (other cases like $a\neq b\neq c\neq d$ or $a=b=c\neq d$ are very esay to check), then the relevant nonzero terms are $\omega _{\ a}^{f}\varepsilon_{fbcd}=\omega _{\ 1}^{0}$ and $\omega _{\ b}^{f}\varepsilon_{afcd}=-\omega _{\ 1}^{0}$. These two terms cancel each other out, so $D\varepsilon _{abcd}=0$.  

\section{Coefficient quantization of ISO(3) Chern-Simons action}
\label{CS}
To construct a Euclidean 3 dimensional gravity theory with the form $e\wedge R$, we need to use the gauge group $ISO(3)$ which is semidirect product of translation group $R(3)$ and rotation group $SO(3)$. Fortunately, the non-compact part $R(3)$ can be contracted to a point, therefore the topological properties of $ISO(3)$ and $SO(3)$ are the same. This can be proved by long exact sequence:

\begin{eqnarray}
&&...\rightarrow \pi_{n}(R3) \rightarrow \pi_{n}(ISO(3)) \rightarrow \pi_{n}(SO(3)) \rightarrow \nonumber\\ &&\pi_{n-1}(R3)\rightarrow 
... \rightarrow \pi_{0}(R3)\rightarrow \pi_{0}(ISO(3)).
\end{eqnarray}

$\pi_{n}(R3)=0$ and $\pi_{n-1}(R3)=0$ lead to $\pi_{n}(ISO(3))=\pi_{n}(SO(3))$. Thus we can use $ISO(3)$ to construct gauge theory and use $SO(3)$ to compute its topological properties.
We define the rotational generators as $J^{ab}$ and the translational generators as $P^{a}$. The non-degenerate invariant bilinear form can be written as $W=\varepsilon _{abc}P^{a}J^{bc}$. So the invariant quadratic form reads:
\begin{equation}
\left\langle P_{a},J_{bc}\right\rangle =\varepsilon _{abc},\ \ %
\left\langle J_{a},J_{b}\right\rangle =\left\langle P_{a},P_{b}\right\rangle
=0. 
\end{equation}

To make $W$ an invariant form ($\left[ P_{a},W\right] =\left[
J_{ab},W\right] =0$, the commutative relations of $ISO(3)$ generator must satisfy:
\begin{equation}
\left[ J_{ab},J_{cd}\right] =\delta _{bc}J_{ad}-\delta _{bd}J_{ac}-\delta
_{ac}J_{bd}+\delta _{ad}J_{bc}, 
\end{equation}
and
\begin{equation}
\left[ P_{a},J_{bc}\right] =\delta _{ab}P_{c}-\delta _{ac}P_{b},\ \ %
\left[ P_{a},P_{b}\right] =0. 
\end{equation}

Let us use these definitions to construct gauge theory for the group
$ISO(3)$. The gauge field is a Lie-algebra-valued one form

\begin{equation}
A_{i}=e_{i}^{a}P_{a}+\frac{1}{2}\omega _{i}^{ab}J_{ab} ,
\end{equation}
Thus, $\delta S_{BR}$ can be written as the usual three-form Chern-Simons action of gauge group $ISO(3)$, where:
\begin{eqnarray}
\int Tr\left( A\wedge dA\right) &=&\frac{1}{2}\int \varepsilon _{abc}\left(
e^{a}\wedge d\omega ^{bc}+\omega ^{ab}\wedge de^{c}\right) \nonumber\\&=&\int
\varepsilon _{abc}e^{a}\wedge d\omega ^{bc} .
\end{eqnarray}
and
\begin{widetext}
\begin{eqnarray}
\frac{2}{3}\int Tr\left( A\wedge A\wedge A\right)&=&\frac{1}{3}\int
Tr\left( [A_{i},A_{j}]dx^{i}\wedge dx^{j}\wedge A\right) \nonumber\\
&=&\frac{1}{3}\int Tr\left[ \left( e^{a}\wedge \omega ^{bc}\left[
P_{a},J_{bc}\right] +\frac{1}{4}\omega ^{ab}\wedge \omega ^{cd}\left[
J_{ab},J_{bd}\right] \right) \wedge A\right] \nonumber\\
&=&\frac{1}{3}\int Tr\left[ \left( 2e^{a}\wedge \omega _{a}^{\ %
c}+\omega _{\ b}^{a}\wedge \omega ^{bc}\right) \wedge A\right] \nonumber\\
&=&\frac{1}{3}\int \left( \varepsilon _{cdf}e^{a}\wedge \omega _{a}^{\
c}\wedge \omega ^{df}+\varepsilon _{acd}e^{d}\wedge \omega _{\ %
b}^{a}\wedge \omega ^{bc}\right).
\end{eqnarray}
\end{widetext}

To simplify this result, we compute the covariant derivative on the scalar:
\begin{eqnarray}
D\left( \varepsilon _{abc}e^{a}\wedge \omega ^{bc}\right) &=&\varepsilon
_{abc}\left( De^{a}\wedge \omega ^{bc}-e^{a}\wedge D\omega ^{bc}\right)
\nonumber\\ &=&\varepsilon _{abc}\left( de^{a}\wedge \omega ^{bc}-e^{a}\wedge d\omega
^{bc}\right) ,
\end{eqnarray}
which gives:
\begin{eqnarray}
\varepsilon _{abc}\omega _{\ d}^{a}\wedge e^{d}\wedge \omega
^{bc}=2\varepsilon _{abc}e^{a}\wedge \omega _{\ d}^{b}\wedge \omega
^{dc} .
\end{eqnarray}
Thus we have:
\begin{eqnarray}
\frac{2}{3}\int Tr\left( A\wedge A\wedge A\right) =\int \varepsilon
_{abc}e^{a}\wedge \omega _{\ d}^{b}\wedge \omega ^{dc} .
\end{eqnarray}
Combining these two equations, we obtain:
\begin{eqnarray}
&&\int Tr\left( A\wedge dA+\frac{2}{3}A\wedge A\wedge A\right) \nonumber\\&=&\int
\varepsilon _{abc}e^{a}\wedge \left( d\omega ^{bc}+\omega _{\ %
d}^{b}\wedge \omega ^{dc}\right) \nonumber\\
&=&\int \varepsilon _{abc}e^{a}\wedge R^{bc}  \nonumber\\
&=&\frac{2\pi }{k_{2}}\delta S'_{top} .
\end{eqnarray}

Now we have proven that $\delta S'_{top}$ can be written as the usual
three-form Chern-Simons action on a three dimensional manifold $M$. Considering that
this three dimensional manifold is the boundary of the four dimensional manifold $W$, it's easy to prove that:
\begin{equation}
\frac{k_{2}}{2\pi }Tr\int_{W}F\wedge F=\frac{k_{2}}{2\pi }Tr\int_{M}\left(
A\wedge dA+\frac{2}{3}A\wedge A\wedge A\right).
\end{equation}
Defining:
\begin{eqnarray}
I_{W}=\frac{k_{2}}{2\pi }Tr\int_{W}F\wedge F ,
\end{eqnarray}
where $F$ is field strength defined as:
\begin{eqnarray}
F_{\mu \nu }=\partial _{\mu }A_{\nu }-\partial _{\nu }A_{\mu }+\left[ A_{\mu
},A_{\nu }\right].
\end{eqnarray}

Now we can choose another different four dimensional space $W'$ whose
boundary is $M$ too. To require these two integral to be equivalent, they must differ by $2n\pi$:
\begin{eqnarray}
I_{W}-I_{W^{\prime }}=I_{X}=2n\pi .
\end{eqnarray}
We can see that in the above equation, the integral region glue together with no boundary. Now on a closed four-manifold $X,$ for $SU(2)$, the quantity $Tr\int_{X}F\wedge F/8\pi^{2}\in \mathbb{Z}$. When $SU(2)$ is replaced by $SO(3)$, which is 2-fold covered by $SU(2)$, we have $Tr\int_{X}F\wedge F/8\pi^{2}\in \mathbb{Z}/4$. Thus:
\begin{equation}
I_{X}=\frac{k_{2}}{2\pi }Tr\int_{X}F\wedge F=m\pi k_{2}=2n\pi .
\end{equation}
with $m$ an arbitrary integer. Finally, we can conclude that $k_{2}=2 \mathbb{Z}$. 

\section{Commutation relations among the constraints}
\label{constraints}

When we compute the commutation relations, it is convenient to consider it in the momentum space. Let us define Fourier transformation as:
\begin{equation}
A(p)=\int d^{4}xe^{ipx}A(x),\ \  A(x)=\int \frac{d^{4}p}{(2\pi)^{4}}e^{-ipx}A(p). 
\end{equation}
For convenience, we denote $\int \frac{d^{4}p}{(2\pi)^{4}}$ as $\int dp$ without causing confuse. The commutation relation in momentum space reads: 
\begin{eqnarray}
[e_{i}^{a}(p),\pi _{b}^{j}(q)]
&=& \int d^{4}x d^{4}y e^{i(px+qy)}[e_{i}^{a}(x),\pi _{b}^{j}(y)]\nonumber\\
&=& i(2\pi)^4\delta _{i}^{j}\delta _{b}^{a}\delta (p+q).
\end{eqnarray}

(1)For $[\mathcal{B}^{ai}(x),\mathcal{C}_{b}(y)]$, we can first calculate the commutator containing one derivative term in momentum space:
\begin{widetext}
\begin{eqnarray}
&&\varepsilon ^{ijk}[\partial_{j}e_{k}^{a}(x),\omega _{b\ l}^{\ f}(y)\pi _{f}^{l}(y)]  \nonumber\\
&=& (-i)\int e^{-i(px+qy+q'y)}dpdq dq' \varepsilon ^{ijk}p_j[e_{k}^{\ a}(p),\omega _{b\ l}^{\ f}(q)\pi _{f}^{l}(q')]\nonumber\\
&=& \int dpdq dq' e^{-i(px+qy-py)} \varepsilon ^{ijk}p_j \omega _{b\ l}^{\ f}(q) (2\pi)^4\delta _{k}^{l}\delta _{f}^{a}\delta (p+q')\nonumber\\
&=& \int dpdq e^{-i(px+qy-py)} \varepsilon ^{ijk}p_j\omega _{b\ k}^{\ a}(q)\nonumber\\
&=& -\int dpdq e^{-i(px+qy)} \varepsilon ^{ijk}p_{j}\omega _{\ bk}^{a}(p+q),
\end{eqnarray}
and 
\begin{eqnarray}
&&\varepsilon ^{ijk}[\omega _{\ fj}^{a}(x)e_{k}^{f}(x),\partial _{l}\pi _{b}^{l}(y)] \nonumber\\
&=& -(-i)\int dpdp'dq  e^{-i(px+p'x+qy)} \varepsilon
^{ijk}q_{l}[\omega _{\ fj}^{a}(p)e_{k}^{f}(p'),\pi _{b}^{l}(q)]
\nonumber\\
&=& \int dpdp'dq e^{-i(px+p'x+qy)}  \varepsilon
^{ijk}q_{l}\omega _{\ fj}^{a}(p)(2\pi)^4 \delta _{k}^{l}\delta _{b}^{f}\delta (p'+q)
\nonumber\\
&=& -\int dpdq  e^{-i(px-qx+qy)} \varepsilon
^{ijk}q_{j}\omega _{\ bk}^{a}(p) 
\nonumber\\
&=& -\int dpdq e^{-i(px+qy)}  \varepsilon
^{ijk}q_{j}\omega _{\ bk}^{a}(p+q).
\end{eqnarray}
The commutator containing two derivative terms is $\varepsilon^{ijk}[\partial _{j}e_{k}^{a}(x),\partial _{l}\pi _{b}^{l}(y)]$. It is vanishing due to $\varepsilon ^{ijk}p_{j}p_{k}=0$. Together with these two terms we have:
\begin{eqnarray}
&&\varepsilon ^{ijk}([\partial _{j}e_{k}^{a}(x),\omega _{b\ j}^{\ f}(y)\pi _{f}^{l}(y)]+[\omega _{\ fj}^{a}(x)e_{k}^{f}(x),\partial _{l}\pi _{b}^{l}(y)]) \nonumber\\ 
&=&-\int dpdq e^{-i(px+qy)}  \varepsilon ^{ijk}(p_{j}+q_{j})\omega _{\ bk}^{a}(p+q)\nonumber\\ 
&=&-\varepsilon ^{ijk}\int dq e^{-iq(y-x)}\int dk e^{-ikx}k_{j}\omega _{\ bk}^{a}(k) \nonumber\\ 
&=&-i\varepsilon ^{ijk}\partial _{j}\omega _{\ bk}^{a}(x)\delta
(y-x).
\end{eqnarray}

The commutator without derivative is trivial:
\begin{equation}
\varepsilon
^{ijk}[\omega _{\ fj}^{a}(x)e_{k}^{f}(x),\omega _{b\ l}^{\ f}(y)\pi _{f}^{l}(y)]=-i \varepsilon ^{ijk}\omega _{\ fj}^{a}\omega _{\ bk}^{f}(x)\delta (x-y).
\end{equation} 
Thus we finally have $\varepsilon ^{ijk}[D_{j}e_{k}^{a}(x),D_{l}\pi _{b}^{l}(y)]=-i \frac{1}{2}\varepsilon ^{ijk}R_{\ bjk}^{a}(x)\delta(x-y)$, or $[\mathcal{B}^{ai}(x),\mathcal{C}_{b}(y)]=-i \mathcal{A}_{\ b}^{a\ i}(x)\delta (x-y)$.
\bigskip 

(2)For $[\mathcal{C}_{a}(x),\mathcal{C}_{b}(y)],$ the commutator containing one derivative reads:
\begin{eqnarray}
&&\lbrack \partial _{i}\pi _{a}^{i}(x),\frac{1}{2}\varepsilon
^{ijk}\varepsilon _{bcdf}R_{ij}^{cd}(y) e_{k}^{f}(y)]+[\frac{1}{2}\varepsilon ^{ijk}\varepsilon _{acdf}R_{ij}^{cd}(x)e_{k}^{f}(x),\partial
_{i}\pi _{b}^{i(y)}] \nonumber\\ 
&=&\int dpdq e^{-i(px+qy)} \frac{1}{2}\varepsilon ^{ijk}\varepsilon_{abcd}(p_{k}+q_{k})R_{ij}^{cd}(p+q) \nonumber\\ 
&=&\frac{1}{2}i \varepsilon ^{ijk}\varepsilon _{abcd}\partial _{i}R_{jk}^{cd}(x)\delta (y-x),
\end{eqnarray}
and the commutator without derivative term yields: 
\begin{eqnarray}
\frac{1}{2}i \varepsilon
^{ijk}\varepsilon _{abcd}\left(\omega _{\ fk}^{c}(x)R_{ij}^{fd}(x)+\omega _{\ fk}^{d}(x)R_{ij}^{cf}(x)\right)\delta(x-y).
\end{eqnarray}
Thus $[\mathcal{C}_{a}(x),\mathcal{C}_{b}(y)]=\frac{1}{2}i \varepsilon ^{ijk}\varepsilon _{abcd}D_{k}R_{ij}^{cd}(x)\delta(x-y)=0$.
The commutator $[\mathcal{D}_{ab}(x),\mathcal{D}_{cd}(y)]$ is similar to this one and we don't show the calculation details again.

\bigskip 
(3)For $[\mathcal{A}^{abi}(x),\mathcal{D}_{cd}(y)],$ the commutator containing one derivative reads:
\begin{eqnarray}
&&\varepsilon ^{ijk}[\partial _{j}\omega _{k}^{ab}(x),\omega _{c\ i}^{\ f}(y)\Pi _{fd}^{i}(y)+\omega _{d\ i}^{\ f}(y)\Pi _{cf}^{i}(y)]+\varepsilon ^{ijk}[\omega _{\ fj}^{a}(x)\omega _{k}^{fb}(x),\partial _{i}\Pi _{cd}^{i}(y)] \nonumber\\ 
&=&-\int dpdq e^{-i(px+qy)} \frac{1}{2}(p_{j}+q_{j})\varepsilon ^{ijk}(\omega _{\ ck}^{a}(p+q)\delta _{d}^{b}-\omega _{\ ck}^{b}(p+q)\delta_{d}^{a}-c\leftrightarrow d) \nonumber\\ 
&=&-\frac{1}{2}i \varepsilon ^{ijk}(\partial _{j}\omega _{\ ck}^{a}(x)\delta _{d}^{b}-\partial _{j}\omega _{\ ck}^{b}(x)\delta _{d}^{a}+\partial _{j}\omega _{\ dk}^{b}(x)\delta _{c}^{a}-\partial _{j}\omega _{\ dk}^{a}(x)\delta _{c}^{b})\delta(x-y)\nonumber\\ 
&=&-2i \varepsilon ^{ijk}\partial _{j}\omega _{\ [dk}^{[b}(x)\delta _{c]}^{a]}\delta(x-y),
\end{eqnarray}
where braket $[]$ means applying the antisymmetrization  only with respect to the internal index, leaving the spacetime index unchanged (always with the total weight 1, e.g., $\omega_{[ab]}=\frac{1}{2}(\omega_{ab}-\omega_{ba})$). 
The commutator without derivative term gives:
\begin{eqnarray}
&&\varepsilon ^{ijk}[\omega _{\ fj}^{a}(x)\omega _{k}^{fb}(x),\omega _{c\ i}^{\ f}(y)\Pi _{fd}^{i}(y)+\omega _{d\ i}^{\ f}(y)\Pi _{cf}^{i}(y)] \nonumber\\ 
&=&i \varepsilon ^{ijk}(\omega _{\ fj}^{a}(x)\omega _{c\ k}^{\ g}(x)\delta _{gd}^{fb}+\omega _{\ fj}^{b}(x)\omega _{c\ k}^{\ g}(x)\delta _{gd}^{af}+\omega _{\ fj}^{a}(x)\omega _{d\ k}^{\ g}(x)\delta _{cg}^{fb}+\omega _{\ fj}^{a}(x)\omega _{d\ k}^{\ g}(x)\delta _{cg}^{af})\delta(x-y) \nonumber\\ 
&=&-\frac{1}{2}i \varepsilon ^{ijk}(\omega _{\ fj}^{b}(x)\omega _{
\ di}^{f}(x)\delta _{c}^{a}-\omega _{\ fj}^{a}(x)\omega _{
\ di}^{f}(x)\delta _{c}^{b}+\omega _{\ fj}^{a}(x)\omega _{
\ ci}^{f}(x)\delta _{d}^{b}-\omega _{\ fj}^{b}\omega _{
\ ci}^{f}(x)\delta _{d}^{a}(x))\delta(x-y) \nonumber\\ 
&=&-2i \varepsilon ^{ijk}\omega _{\ fj}^{[b}(x)\omega _{\ [di}^{|f|}(x)\delta _{c]}^{a]}\delta(x-y).
\end{eqnarray}
Together with the above two terms, we have $[\mathcal{A}^{abi}(x),\mathcal{D}_{cd}(y)]=-i \varepsilon^{ijk}\delta _{[c}^{[a} R_{\ d]jk}^{b]}(x)\delta(x-y)=-2i \delta _{[c}^{[a}\mathcal{A}_{\ d]}^{b]\ i}(x)\delta(x-y)$.

\bigskip
(4) For $[\mathcal{B}^{ai}(x),\mathcal{D}_{cd}(y)],$ the commutator containing one derivative reads:
\begin{eqnarray}
&&\lbrack \varepsilon ^{ijk}\omega _{\ fj}^{a}(x)e_{k}^{f}(x),\partial _{i}\Pi _{cd}^{i}(y)]+[\varepsilon ^{ijk}\partial _{j}e_{k}^{a}(x),\frac{1}{2}(\pi _{c}^{i}(y)e_{di}(y)-\pi _{d}^{i}(y)e_{ci}(y))] \nonumber\\ 
&=&\int dpdq e^{-i(px+qy)} \varepsilon ^{ijk}[q_{j}e_{fk}\delta _{cd}^{af}+\frac{1}{2}p_{j}(e_{dk}\delta _{c}^{a}-e_{ck}\delta _{d}^{a})] \nonumber\\ 
&=&\int dpdq e^{-i(px+qy)}\frac{1}{2}\varepsilon ^{ijk}(p_{j}+q_{j})(e_{dk}\delta_{c}^{a}-e_{ck}\delta _{d}^{a}) \nonumber\\ 
&=&i \varepsilon ^{ijk}\partial _{j}e_{[dk}(x)\delta _{c]}^{a}\delta(x-y).
\end{eqnarray}
and the commutator without derivative term gives:
\begin{eqnarray}
&&\lbrack \varepsilon ^{ijk}\omega _{j}^{af}(x)e_{fk}(x),\omega _{c\ i}^{\ g}(y)\Pi _{gd}^{i}(y)+\omega _{d\ i}^{\ g}(y)\Pi _{cg}^{i}(y)]+[\varepsilon ^{ijk}\omega _{j}^{af}(x)e_{fk}(x),\frac{1}{2}(\pi _{
c}^{i}(y)e_{di}(y)-\pi _{d}^{i}(y)e_{ci}(y))] \nonumber\\ 
&=&i \varepsilon ^{ijk}[\omega _{c\ j}^{\ g}(x)e_{fk}(x)\delta _{gd}^{af}+\omega_{d\ j}^{\ g}(x)e_{fk}(x)\delta _{cg}^{af}+\frac{1}{2}(\omega _{\ cj}^{a}(x)e_{dk}(x)-\omega _{\ dj}^{a}(x)e_{ck}(x))] \delta(x-y)\nonumber\\ 
&=&\frac{1}{2}i \varepsilon ^{ijk}(\omega _{d\ j}^{\ f}(x)e_{fk}(x)\delta
_{c}^{a}-\omega _{c\ j}^{\ f}(x)e_{fk}(x)\delta _{d}^{a})\delta(x-y).
\end{eqnarray}
Together with these two terms, we have $[\mathcal{B}^{ai}(x),\mathcal{D}_{cd}(y)]=i\varepsilon ^{ijk} D_{j}e_{[dk}(x)\delta^{a}_{c]}\delta(x-y)=i \delta _{[c}^{a}\mathcal{B}_{d]}^{i}(x)\delta(x-y).$

\bigskip

(5)For $[\mathcal{C}^{a}(x),\mathcal{D}_{cd}(y)],$ the commutator $[D_{i}\pi _{a}^{i}(x),\mathcal{D}_{cd}(y)]$ 
is similar to $[\mathcal{B}^{ai}(x),\mathcal{D}_{cd}(y)]$, so we have $[D_{i}\pi _{a}^{i}(x),\mathcal{D}_{cd}(y)]=i D_{i}\pi _{[d}^{i}(x)\delta _{c]}^{a}\delta(x-y).$ Next we need to
compute the commutator $[\frac{1}{2}\varepsilon ^{ijk}\varepsilon _{abcd} R_{ij}^{bc}(x)e_{k}^{d}(x),D_{i}\Pi _{cd}^{i}(y)]$ (for the sake of indices, we rewrite it as $[\frac{1}{2}\varepsilon ^{ijk}\varepsilon _{abmn}e_{i}^{b}(x)R_{jk}^{mn}(x),D_{i}\Pi _{cd}^{i}(y)]$). This term is similar to $[\mathcal{A}^{abi}(x),\mathcal{D}_{cd}(y)]$ except for an extra variable $e_{i}^{b}$. The calculation of commutators without derivative is trivial. We focus on the commutator containing one derivative, which gives rise to:

\begin{eqnarray}
&&\lbrack \varepsilon ^{ijk}\varepsilon _{abmn}e_{i}^{b}(x)\partial
_{j}\omega _{k}^{mn}(x),\omega _{c\ i}^{\ g}(y)\Pi _{gd}^{i}(y)+\omega _{d\ i}^{\ g}(y)\Pi _{cg}^{i}(y)]+
[\varepsilon^{ijk}\varepsilon _{abmn}e_{i}^{b}(x)\omega _{\ fj}^{m}(x)\omega _{k}^{fn}(x),\partial _{i}\Pi _{cd}^{i}(y)] \nonumber\\ 
&=&(-i)\int dpdq e^{-i(px+qy)} \varepsilon ^{ijk}\varepsilon _{abmn}[\int dp_{1}e_{i}^{b}(p-p_{1})p_{1j}\omega _{k}^{mn}(p_{1}),\int dq_{1}(\omega _{c\ i}^{\ g}(q-q_{1})\Pi _{gd}^{i}(q_{1})+\omega _{d\ i}^{\ g}(q-q_{1})\Pi _{cd}^{i}(q_{1})] \nonumber\\ 
&&+(-i)\int dpdq e^{-i(px+qy)} \varepsilon ^{ijk}\varepsilon _{abmn}[\int dp_{1}dp_{2}e_{i}^{b}(p-p_{1}-p_{2})\omega _{\ fj}^{m}(p_{1})\omega _{k}^{fn}(p_{2}),q_{i}\Pi _{cd}^{i}(q)] \nonumber\\ 
&=&\int dpdq e^{-i(px+qy)} \varepsilon ^{ijk}\varepsilon _{abmn}\{\int dp_{1}e_{i}^{b}(p-p_{1})p_{1j}[\omega _{c\ k}^{\ g}(q+p_{1})\delta
_{gd}^{mn}+\omega _{d\ k}^{\ g}(q+p_{1})\delta _{cg}^{mn}] \nonumber\\ 
&&+\int dp_{2}e_{i}^{b}(p+q-p_{2})q_{j}[\omega _{f\ k}^{\ 
n}(p_{2})\delta _{cd}^{mf}+\omega _{f\ k}^{\ m}(p_{2})\delta
_{cd}^{fn}]\} \nonumber\\ 
&=&\int dpdq e^{-i(px+qy)} \varepsilon ^{ijk}\varepsilon _{abmn}\int dp_{1}e_{i}^{b}(p-p_{1})\times (p_{1j}+q_{j})[\omega _{c\ k}^{\ m}(q+p_{1})\delta
_{d}^{n}+\omega _{d\ k}^{\ n}(q+p_{1})\delta _{c}^{m}] \nonumber\\ 
&=&i \varepsilon ^{ijk}\varepsilon _{abmn}e_{i}^{b}(x)\left(\partial _{j}\omega _{c\ k}^{\ m}(x)\delta _{d}^{n}+\partial _{j}\omega _{d\ k}^{\ n}(x)\delta_{c}^{m}\right)\delta(x-y),
\end{eqnarray}
Thus we have:
\begin{eqnarray}
[\frac{1}{2}\varepsilon ^{ijk}\varepsilon _{abcd}e_{i}^{b}(x)R_{jk}^{mn}(x),D_{i}\Pi _{cd}^{i}(y)]=\frac{1}{2}i \varepsilon
^{ijk}\varepsilon _{abmn}e_{i}^{b}(x)\left(R_{c\ jk}^{\ m}(x)\delta _{d}^{n}+R_{d\ jk}^{\ n}(x)\delta _{c}^{m}\right)\delta(x-y).
\end{eqnarray}
Together with: 
\begin{eqnarray}
[\frac{1}{2}\varepsilon ^{ijk}\varepsilon
_{abcd}e_{i}^{b}(x)R_{jk}^{cd}(x),\frac{1}{2}(\pi _{c}^{i}(y)e_{di}(y)-\pi _{d}^{i}(y)e_{ci}(y))]=\frac{1}{4}i \varepsilon ^{ijk}R_{jk}^{mn}(x)(e_{di}(x)\varepsilon _{acmn}-e_{ci}(x)\varepsilon _{admn})\delta(x-y),
\end{eqnarray}
we obtain:
\begin{eqnarray}
\lbrack \frac{1}{2}\varepsilon ^{ijk}\varepsilon _{abmn}e_{i}^{b}(x)R_{jk}^{mn}(x),\mathcal{D}_{cd}(y)] &=&\frac{1}{2}i \varepsilon ^{ijk}\varepsilon _{abmn}e_{i}^{b}(x)(R_{c\ jk}^{\ m}(x)\delta _{d}^{n}+R_{d\ jk}^{\ n}(x)\delta _{c}^{m})\delta(x-y)\nonumber\\ 
&& +\frac{1}{4}i \varepsilon ^{ijk}R_{jk}^{mn}(x)(e_{di}(x)\varepsilon
_{acmn}-e_{ci}(x)\varepsilon _{admn})\delta(x-y) \nonumber\\ 
&=&\frac{1}{4}i \varepsilon ^{ijk}e_{i}^{b}(x)R_{jk}^{mn}(x)(\eta
_{ac}\varepsilon _{dbmn}-\eta _{ad}\varepsilon _{cbmn})\delta(x-y).
\end{eqnarray}
\end{widetext}
The second equation can be proven by checking their components in two cases: $a=c\neq d$  and $a\neq c\neq d$. We ignore the spacetime index $i,j,k$ here. Without loss of generality, we consider:

(1) $a=c=0,$ $d=1.$ The nonzero component of the first line is $\frac{i}{2}(e^{2}R^{\ 3}_{0}-e^{3}R^{\ 2}_{0})-\frac{i}{2}e_{0}R^{23},$ and the nonzero component of the second line is $\frac{i}{2}(e^{2}R^{30}+e^{3}R^{02}+e^{0}R^{23})$. They are the same if we raise up the index 0.

(2) $a=2,$ $c=0,$ $d=1.$ The nonzero component of the first line is $\frac{i}{2}(-R_{1}^{\ 3}e^{1}-R_{0}^{\ 3}e^{0})+\frac{i}{2}(R_{1}^{\ 3}e^{1}+R_{0}^{\ 3}e^{0})=0;$. The second line is zero apparently.
\begin{widetext}
Thus we finally prove:
\begin{eqnarray}
[\mathcal{C}^{a}(x),\mathcal{D}_{cd}(y)]&=&i \left(D_{i}\pi _{[d}^{i}(x)\delta _{c]}^{a}+\frac{1}{2}i \delta _{\lbrack c}^{[a}\varepsilon
_{d]bmn}\varepsilon ^{ijk}e_{i}^{b}(x)R_{jk}^{mn}(x)\right)\delta(x-y)\nonumber\\
&=&i\left( D_{i}\pi _{[d}^{i}(x)\delta _{c]}^{a}+\frac{1}{2}i \delta _{\lbrack c}^{[a}\varepsilon
_{d]mnb}\varepsilon^{ijk}R_{ij}^{mn}(x)e_{k}^{b}(x)\right)\delta(x-y)\nonumber\\ 
&=&i \delta
_{[c}^{a} \mathcal{C}_{d]}(x)\delta(x-y).
\end{eqnarray}
\end{widetext}

\section{The explicit solution of $\widetilde{B}$}\label{calofB}

Assuming $\widetilde{B}_{d}=\frac{l_{p}}{\kappa}
\varepsilon _{abcd} e^{b} \wedge K^{cd}+
\frac{l_{p}}{2}\varepsilon _{abcd} e^{b}\wedge e^{f}\widetilde{T}_{f}^{\ cd}$. We can prove that this solution satisfy Eq. (\ref{BwithM}), which is:
\begin{equation}
\frac{1}{\kappa}T^{c}\wedge e^{d}\varepsilon _{abcd}+\frac{1}{2l_{p}}\left( \widetilde{B}_{a}\wedge e_{b}-\widetilde{B}_{b}\wedge e_{a}\right)=\widetilde{T}_{ab}.
\end{equation}
First we prove that:
\begin{widetext}
\begin{equation}
\frac{1}{2}\left(\varepsilon _{acmn} e^{c}\wedge K^{mn}\wedge 
e_{b}-\varepsilon _{bcmn} e^{c}\wedge K^{mn}\wedge e_{a}\right) =-T^{c}\wedge e^{d}\varepsilon _{abcd} ,
\end{equation}
or equivalently:
\begin{equation}
\frac{1}{2}\left( \varepsilon _{acmn}e_{b}-\varepsilon
_{bcmn}e_{a}\right) \wedge K^{mn}\wedge e^{c}=K_{\ f}^{c}\wedge
e^{f}\wedge e^{d}\varepsilon _{abcd} .
\end{equation}
this can be verified by considering its components. Without loss of generality we set $a=0,$ $b=1$. The nonzero component of left hand side is
\begin{equation}
e_{1}\wedge \left( K^{12}\wedge e^{3}+K^{31}\wedge e^{2}\right) +e_{0}\wedge
\left( K^{02}\wedge e^{3}+K^{30}\wedge e^{2}\right),
\end{equation}
where we have used $e_{i}\wedge e_{i}=0$. The nonzero component of right hand side is
\begin{eqnarray}
&&\left( K_{\ 0}^{2}\wedge e^{0}\wedge e^{3}+K_{\ 1}^{2}\wedge
e^{1}\wedge e^{3}\right)-\left( K_{\ 0}^{3}\wedge e^{0}\wedge
e^{2}+K_{\ 1}^{3}\wedge e^{1}\wedge e^{2}\right) \nonumber\\
&=&e_{1}\wedge \left( K^{12}\wedge e^{3}+K^{31}\wedge e^{2}\right)
+e_{0}\wedge \left( K^{02}\wedge e^{3}+K^{30}\wedge e^{2}\right),
\end{eqnarray}
so the left hand side is equal to right hand side.

Then we prove:
\begin{equation}
\frac{1}{4}\widetilde{T}_{f}^{\ mn}\left( \varepsilon _{acmn}e^{c}\wedge e^{f}\wedge e_{b}-\varepsilon _{bcmn} e^{c}\wedge e^{f}\wedge e_{a}\right) =\widetilde{T}_{ab},
\end{equation}
or equivalently:
\begin{equation}
-\frac{1}{4}\widetilde{T}_{f}^{\ mn}e^{f}\wedge
e^{c}\wedge \left( \varepsilon _{acmn}e_{b}-\varepsilon _{bcmn}e_{a}\right)
=\frac{1}{6}\widetilde{T}_{\ ab}^{g}\varepsilon _{gcdf}e^{c}\wedge
e^{d}\wedge e^{f} .
\end{equation}
We can also verify it by considering the components. Setting $a=0,$ $b=1$, the nonzero component left hand side is:
\begin{eqnarray}
&&-\frac{1}{2}\left( \widetilde{T}_{0}^{\ 12}e^{0}\wedge e^{3}+\widetilde{T}_{0}^{\ 31}e^{0}\wedge e^{2}\right) \wedge e_{1}-\frac{1}{2}\left( \widetilde{T}_{1}^{\ 02}e^{1}\wedge e^{3}+\widetilde{T}_{1}^{\ 30}e^{1}\wedge
e^{2}\right) \wedge e_{0} \nonumber\\
&=&\widetilde{T}^{012}e_{0}\wedge e_{1}\wedge e^{3}-\widetilde{T}^{013}e_{0}\wedge e_{1}\wedge e^{2},
\end{eqnarray}
\end{widetext}
and the nonzero component of right hand side is:
\begin{eqnarray}
&&-\widetilde{T}_{\ 01}^{3}e^{0}\wedge e^{1}\wedge e^{2}+\widetilde{T}_{\ 01}^{2}e^{0}\wedge e^{1}\wedge e^{3} \nonumber\\
&=&\widetilde{T}^{012}e_{0}\wedge e_{1}\wedge e^{3}-\widetilde{T}^{013}e_{0}\wedge e_{1}\wedge e^{2},
\end{eqnarray}
Therefore, the left hand side is equal to right hand side. Combining the above two parts, we finaly prove:
\begin{equation}
\widetilde{B}_{d}=\frac{l_{p}}{\kappa}
\varepsilon _{abcd} e^{b} \wedge K^{cd}+
\frac{l_{p}}{2}\varepsilon _{abcd} e^{b}\wedge e^{f}\widetilde{T}_{f}^{\ cd}.
\end{equation}
Without Dirac field, $\widetilde{B}_{d}$ has a very simple solution:
\begin{equation}
\widetilde{B}_{d}=\frac{l_{p}}{\kappa}
\varepsilon _{abcd} e^{b} \wedge K^{cd}.
\end{equation}
In order to solve the $\widetilde{B}_{d}$ for the most general TQFT action(without Dirac field):
\begin{equation}
\frac{1}{\kappa} T^{c}\wedge e^{d}\varepsilon _{abcd}+(\frac{1}{l_{p}} \widetilde{B}+(\psi -\lambda
)T)_{[a}\wedge e_{b]}=0,
\end{equation}
we can just regard $\widetilde{B}'_{d}=\widetilde{B}_{d}+l_{p}(\psi -\lambda
)T_{a}$ as the new variable. Then it becomes the same equation as the previous case without Dirac field, and we have:
\begin{equation}
\widetilde{B}'_{d}=\frac{l_{p}}{\kappa} \varepsilon _{abcd} e^{b} \wedge K^{cd} \Rightarrow \widetilde{B}_{d}=\frac{l_{p}}{\kappa}\varepsilon _{abcd} e^{b} \wedge K^{cd}-l_{p}(\psi -\lambda)T_{a}.
\end{equation}

\begin{widetext}
\section{EOM and energy-momentum tensor of the Dirac field}
\label{deriofphi}

To compute the variation with respect to $\overline{\psi }$, which gives the EOM of the Dirac field, we foucs on the hermitian conjugate term in Eq.(\ref{Sdirac2}):

\begin{eqnarray}
&&\frac{1}{3!}\int \varepsilon_{abcd} e^{a}\wedge e^{b}\wedge e^{c}\wedge \left(\frac{1}{2}\overline{\psi}\gamma
^{d} \nabla\psi\right)^{\dagger}\nonumber\\
&=&\frac{1}{3!}\frac{1}{2}\int \varepsilon_{abcd} e^{a}\wedge e^{b}\wedge e^{c}\wedge \left(- d \overline{\psi}\gamma^{d} \psi
+\omega_{fg}\overline{\psi }\Pi^{fg}\gamma^{d}\psi\right)\nonumber\\
&=&-\frac{1}{2!}\frac{1}{2}\int \varepsilon_{abcd}\ de^{a}\wedge e^{b}\wedge e^{c}\overline{\psi}\gamma^{d} \psi+\frac{1}{3!}\frac{1}{2}\int \varepsilon_{abcd} e^{a}\wedge e^{b}\wedge e^{c}\wedge  \overline{\psi}\gamma^{d} d\psi\nonumber\\
&&+\frac{1}{3!}\frac{i}{2}\int \varepsilon_{abcd} e^{a}\wedge e^{b}\wedge e^{c}\wedge  \omega_{fg}\overline{\psi }\left(\gamma^{d}\Pi^{fg}+[\Pi^{fg},\gamma^{d}]\right)\psi \nonumber\\
&=&\frac{1}{3!}\frac{1}{2}\int \varepsilon_{abcd} e^{a}\wedge e^{b}\wedge e^{c}\wedge \overline{\psi}\gamma
^{d} \nabla\psi -\frac{1}{2!}\frac{1}{2}\int \varepsilon_{abcd}\ de^{a}\wedge e^{b}\wedge e^{c}\overline{\psi}\gamma^{d} \psi-\frac{1}{3!}\frac{1}{2}\int \varepsilon_{abcd} e^{a}\wedge e^{b}\wedge e^{c}\wedge \omega_{\ f}^{d}\overline{\psi}\gamma ^{f} \psi\nonumber\\
&=&\frac{1}{3!}\frac{1}{2}\int \varepsilon_{abcd} e^{a}\wedge e^{b}\wedge e^{c}\wedge \overline{\psi}\gamma
^{d} \nabla\psi -\frac{1}{2!}\frac{1}{2}\int \varepsilon_{abcd}\ de^{a}\wedge e^{b}\wedge e^{c}\overline{\psi}\gamma^{d} \psi-\frac{1}{2!}\frac{1}{2}\int \varepsilon_{abcd} \omega_{\ f}^{a} \wedge e^{f}\wedge e^{b}\wedge e^{c}\overline{\psi}\gamma ^{d} \psi\nonumber\\
&=&\frac{1}{3!}\frac{1}{2}\int \varepsilon_{abcd} e^{a}\wedge e^{b}\wedge e^{c}\wedge \overline{\psi}\gamma
^{d} \nabla\psi -\frac{1}{2!}\frac{1}{2}\int \varepsilon_{abcd}T^{a}\wedge e^{b}\wedge e^{c}\overline{\psi}\gamma^{d}\psi.
\end{eqnarray}
In the third step we use the following identity:

\begin{eqnarray}
\omega _{fg}\left[ \Pi ^{fg},\gamma ^{d}\right] &=&\frac{1}{8
}\omega _{fg}\left( \gamma ^{f}\gamma ^{g}\gamma ^{d}-\gamma ^{g}\gamma
^{f}\gamma ^{d}-\gamma ^{d}\gamma ^{f}\gamma ^{g}+\gamma ^{d}\gamma
^{g}\gamma ^{f}\right) \nonumber\\
&=&\frac{1}{8}\omega _{fg}\left( 2\eta ^{gd}\gamma ^{f}-2\eta ^{fd}\gamma^{g}+2\eta ^{gd}\gamma ^{f}-2\eta ^{fd}\gamma ^{g}\right) \nonumber\\
&=&\frac{1}{2}\omega _{fg}\left( \eta ^{gd}\gamma ^{f}-\eta ^{fd}\gamma^{g}\right) \nonumber\\
&=&-\omega _{\ f}^{d}\gamma ^{f},
\end{eqnarray}
where in the last step we use the identity:
\begin{equation}
\varepsilon_{abcd} \omega _{\ f}^{d}=-\varepsilon_{dbcf} \omega _{\ a}^{d}-\varepsilon_{adcf} \omega _{\ b}^{d}-\varepsilon_{abdf} \omega _{\ c}^{d}.
\end{equation}
Thus the action can be rewritten as:
\begin{equation}
S_{Dirac}=\frac{1}{3!}\int \varepsilon_{abcd} e^{a}\wedge e^{b}\wedge e^{c}\wedge \overline{\psi}\gamma
^{d} \nabla\psi + \frac{1}{4!}\int \varepsilon_{abcd} e^{a}\wedge e^{b}\wedge e^{c}\wedge e^{d} m\overline{\psi}\psi-\frac{1}{2!}\frac{1}{2}\int \varepsilon_{abcd}T^{a}\wedge e^{b}\wedge e^{c}\overline{\psi}\gamma^{d}\psi.
\end{equation}
Variation with respect to $\overline{\psi }$ gives:
\begin{equation}
\frac{1}{3!}  \varepsilon_{abcd} e^{a}\wedge e^{b}\wedge e^{c}\wedge \gamma ^{d} \nabla\psi 
+\frac{1}{4!} \varepsilon_{abcd} e^{a}\wedge e^{b}\wedge e^{c}\wedge e^{d} m\psi -\frac{1}{4}\varepsilon_{abcd}T^{a}\wedge e^{b}\wedge e^{c} \gamma^{d}\psi =0.
\end{equation}

\section{The antisymmetric part of generalized Einstein equaiton with Dirac field}\label{antisymmetric}

The antisymmetric part of Eq. (\ref{EOMDirac}) reads:
\begin{eqnarray}
\frac{1}{2\pi}\left( \Sigma _{fcb}\Sigma _{\ \ d}^{cb}-\Sigma
_{dcb}\Sigma _{\ \ f}^{cb}\right)&& -\frac{1}{2\pi}\sigma _{c}\left( \Sigma
_{\ fd}^{c}-\Sigma _{\ df}^{c}\right) +\frac{1}{4}l_{p}\left( 
\widetilde{T}_{fab}\Sigma _{d}^{\ ab}-\widetilde{T}_{dab}\Sigma _{f}^{\ ab}\right) \nonumber\\
&&-\frac{1}{\pi}\sigma _{a}\Sigma _{fd}^{\ \ a}+\frac{1}{2}\widetilde{D}_{i}\widetilde{T}_{\
 fd}^{i}+\frac{1}{2}t_{[fd]}=0 ,
\end{eqnarray}
or equivalently
\begin{equation}
\frac{1}{4}\left( \sigma _{fcb}\sigma _{\ \ d}^{cb}-\sigma
_{dcb}\sigma _{\ \ f}^{cb}\right) +\frac{1}{2}\sigma _{a}\left( \sigma
_{fd}^{\ \ a}-\sigma _{df}^{\ \ a}\right) =-\frac{\pi}{2}\widetilde{D}_{i}\widetilde{T}_{\ fd}^{i}-\frac{\pi}{2}t_{[fd]}-\frac{\pi}{4} l_{p}\left( \widetilde{T}_{fab}\Sigma _{d}^{\ ab}-\widetilde{T}_{dab}\Sigma _{f}^{\ ab}\right), \label{antiequation}
\end{equation}
where  $\Sigma _{aij}=-\frac{1}{2}\left( \sigma _{aij}+\sigma _{ija}-\sigma _{jai}\right)$ is the contorsion of loop source, $t_{\ d}^{f}=-\frac{1}{2}\overline{\psi }\gamma ^{f}e_{d}^{\mu }\left(
\partial _{\mu }+\Omega _{\mu }\right) \psi +\left( h.c.\right) $ is the energy momentum tensor of Dirac field, and $\Omega _{\mu}=\omega _{\ \ \mu}^{ab}\Pi _{ab}$. The Dirac current is defined as $\widetilde{T}^{abc}=-\frac{1}{2}
\overline{\psi }\left( \gamma ^{a}\Pi ^{bc}+\Pi ^{bc}\gamma ^{a}\right) \psi 
$. In the following we will prove that by using the EOM of the Dirac field:
\begin{equation}
\gamma ^{a}e_{a}^{\mu }\left( \partial _{\mu }+\Omega _{\mu }\right) \psi
+m\psi -\frac{1}{2}T_{\ ab}^{a}\gamma ^{b}\psi =0,
\end{equation}
we can simplify the expression $\widetilde{D}_{i}\widetilde{T}_{\ fd}^{i}+t_{[fd]}=-\frac{1}{2}(T_{\ ab}^{f}\widetilde{T}^{abd}-T_{\ ab}^{d}\widetilde{T}^{abf})$. For convenience, we define $T_{\ ab}^{a}\equiv T_{b}$.

First we try to express $\widetilde{D}_{i}\widetilde{T}^{ifd}$ as:
\begin{equation}
\widetilde{D}_{i}\widetilde{T}^{ifd}=-\frac{1}{2}\partial _{i}[\overline{\psi} \left( \gamma ^{i}\Pi ^{fd}+\Pi ^{fd}\gamma ^{i}\right) \psi] +\Gamma_{\ bi}^{i}\widetilde{T}^{bfd}+
\Gamma_{\ bi}^{f}\widetilde{T}^{ibd}+\Gamma_{\ bi}^{d} \widetilde{T}^{ifb}. \label{DTdirac}
\end{equation}
Applying the commutation relation between gamma matrices, we obtain:
\begin{equation}
\gamma ^{i}\Pi ^{fd}=\Pi ^{fd}\gamma ^{i}+\frac{1}{2}(\eta ^{fi}\gamma
^{d}-\eta ^{di}\gamma ^{f}), \label{omegapi}
\end{equation}
such that Eq. (\ref{DTdirac}) becomes:
\begin{eqnarray}
\widetilde{D}_{i}\widetilde{T}^{ifd}&=&-(\partial _{i}\overline{\psi })\gamma ^{i}\Pi ^{fd}\psi -\overline{\psi } \Pi ^{fd}\gamma ^{i}(\partial _{i}\psi )+\Gamma_{\ bi}^{i}\widetilde{T}^{bfd}+
\Gamma_{\ bi}^{f}\widetilde{T}^{ibd}+\Gamma_{\ bi}^{d} \widetilde{T}^{ifb}\nonumber\\
&&-\frac{1}{4}(\partial ^{d}\overline{\psi }\gamma ^{f}\psi -\partial ^{f}\overline{\psi }\gamma ^{d}\psi +\overline{\psi }\gamma ^{d}\partial ^{f}\psi -\overline{\psi }\gamma ^{f}\partial ^{d}\psi ). \label{derivative}
\end{eqnarray}
Actually the second line of Eq. (\ref{derivative}) will cancel out the partial derivative terms of $t_{[fd]}$. Therefore, we can just focus on the first line. Applying the EOM of Dirac field, we get:
\begin{equation}
(\partial _{i}\overline{\psi })\gamma ^{i}\Pi ^{fd}\psi =\overline{\psi }(\Omega _{i}\gamma ^{i}+m+\frac{1}{2}T_{i}\gamma ^{i})\Pi ^{fd}\psi,
\end{equation}
and
\begin{equation}
\overline{\psi }\Pi ^{fd}\gamma ^{i}(\partial _{i}\psi )=\overline{\psi }\Pi
^{fd}(-\gamma ^{i}\Omega _{i}-m+\frac{1}{2}T_{i}\gamma ^{i})\psi .
\end{equation}
Combining them and using the definition $\widetilde{T}^{abc}=-\frac{1}{2}
\overline{\psi }\left( \gamma ^{a}\Pi ^{bc}+\Pi ^{bc}\gamma ^{a}\right) \psi
,$ we obtain:
\begin{eqnarray}
\widetilde{D}_{i}\widetilde{T}^{ifd}&=&-\overline{\psi }(\Omega _{i}\gamma^{i}\Pi ^{fd}-\Pi ^{fd}\gamma ^{i}\Omega _{i})\psi +\omega _{\ bi}^{i}\widetilde{T}^{bfd}+\Gamma_{\ bi}^{f}\widetilde{T}^{ibd}+\Gamma_{\ bi}^{d}\widetilde{T}^{ifb}\nonumber\\
&&-\frac{1}{4}(\partial ^{d}\overline{\psi }\gamma ^{f}\psi -\partial ^{f}\overline{\psi }\gamma ^{d}\psi +\overline{\psi }\gamma ^{d}\partial ^{f}\psi -\overline{\psi }\gamma ^{f}\partial ^{d}\psi )\nonumber\\
&=&-\frac{1}{2}\overline{\psi }(\Omega ^{f}\gamma ^{d}-\Omega
^{d}\gamma ^{f})\psi-\overline{\psi }(\omega _{\ ai}^{f}\Pi ^{da}\gamma ^{i}-\omega _{\  ai}^{d}\Pi ^{fa}\gamma ^{i}+\Pi ^{fd}\gamma ^{a}\omega _{ab}^{\ 
\ b})\psi +\omega _{\ bi}^{i}\widetilde{T}^{bfd}+\Gamma_{\ bi}^{f}\widetilde{T}^{ibd}+\Gamma_{\ bi}^{d}\widetilde{T}^{ifb}\nonumber\\
&&-\frac{1}{4}(\partial ^{d}\overline{\psi }\gamma ^{f}\psi -\partial ^{f}\overline{\psi }\gamma ^{d}\psi +\overline{\psi }\gamma ^{d}\partial ^{f}\psi -\overline{\psi }\gamma ^{f}\partial ^{d}\psi ),\nonumber\\  
 \label{DTSimple}
\end{eqnarray}
where we have applied the relation $T_{b}+\Gamma_{\ bi}^{i}=\omega _{\ bi}^{i}$ and the identity:

\begin{eqnarray}
\Omega _{i}\gamma ^{i}\Pi ^{fd} &=&\Pi ^{fd}\gamma ^{i}\Omega _{i}+\frac{1}{2%
}\Omega _{i}(\eta ^{fi}\gamma ^{d}-\eta ^{di}\gamma ^{f}) +\frac{1}{2}\omega _{abi}(\eta ^{bd}\Pi ^{fa}+\eta ^{bf}\Pi ^{ad}+\eta
^{ad}\Pi ^{bf}+\eta ^{af}\Pi ^{db})\gamma ^{i}  \nonumber +\frac{1}{2}\Pi ^{fd}(\eta ^{bi}\gamma ^{a}-\eta ^{ai}\gamma ^{b})\omega
_{abi}  \nonumber\\ 
&=&\Pi ^{fd}\gamma ^{i}\Omega _{i}+\frac{1}{2}(\Omega ^{f}\gamma ^{d}-\Omega
^{d}\gamma ^{f})+\omega _{\ ai}^{f}\Pi ^{da}\gamma ^{i}-\omega _{\ ai}^{d}\Pi
^{fa}\gamma ^{i}+\Pi ^{fd}\gamma ^{a}\omega _{ab}^{\ \ b}.
\end{eqnarray}

Next we consider $t_{[fd]}$. The explicit form of $t_{[fd]}$ is:
\begin{eqnarray}
t_{[fd]}&=&\frac{1}{4}(\partial _{d}\overline{\psi }\gamma _{f}\psi -\partial _{f}\overline{\psi }\gamma_{d}\psi +\overline{\psi }\gamma_{d}\partial _{f}\psi -\overline{\psi }\gamma_{f}\partial_{d}\psi )-\frac{1}{4}\overline{\psi }(\gamma _{f}\Omega _{d}+\Omega
_{d}\gamma _{f}-\gamma _{d}\Omega _{f}-\Omega _{f}\gamma _{d})\psi. 
\end{eqnarray}
Adding them together, we obtain:
\begin{eqnarray}
\widetilde{D}_{i}\widetilde{T}^{ifd}+t^{[fd]} &=&-\frac{1}{4}\overline{\psi }
(\Omega ^{f}\gamma ^{d}-\Omega ^{d}\gamma ^{f}+\gamma ^{f}\Omega ^{d}-\gamma
^{d}\Omega ^{f})\psi-\overline{\psi }(\omega _{\ ai}^{f}\Pi ^{da}\gamma ^{i}-\omega _{\  ai}^{d}\Pi ^{fa}\gamma ^{i}+\Pi ^{fd}\gamma ^{a}\omega _{ab}^{\ 
\ b})\psi  \nonumber \\
&&+\omega _{\ bi}^{i}\widetilde{T}^{bfd}+\Gamma_{\ bi}^{f}
\widetilde{T}^{ibd}+\Gamma_{\ bi}^{d}\widetilde{T}^{ifb}. \label{DT-t}
\end{eqnarray}
According to commutation relation in Eq. (\ref{omegapi}), the first term in Eq. (\ref{DT-t}) can be expressed as:
\begin{eqnarray}
-\frac{1}{4}\overline{\psi }(\Omega ^{f}\gamma ^{d}-\Omega ^{d}\gamma
^{f}+\gamma ^{f}\Omega ^{d}-\gamma ^{d}\Omega ^{f})\psi &=&-\frac{1}{8}\overline{\psi}(\eta ^{af}\gamma ^{b}-\eta ^{bf}\gamma ^{a})\omega _{ab}^{\ \ d}\psi+\frac{1}{8}\overline{\psi }(\eta ^{ad}\gamma ^{b}-\eta ^{bd}\gamma^{a})\omega _{ab}^{\ \ f}\psi \nonumber\\
&=&\frac{1}{4}\overline{\psi }(\omega _{i}^{\ fd}\gamma ^{i}-\omega
_{i}^{\ df}\gamma ^{i})\psi,
\end{eqnarray}
and the second term in Eq. (\ref{DT-t}) can be expressed as:
\begin{equation}
-\overline{\psi }(\omega _{\ ai}^{f}\Pi ^{da}\gamma ^{i}-\omega _{\ ai}^{d}\Pi ^{fa}\gamma ^{i}+\Pi ^{fd}\gamma ^{a}\omega _{ab}^{\ 
\ b})\psi =\omega _{\ ab}^{f}\widetilde{T}^{abd}-\omega _{\ ab}^{d}\widetilde{T}^{abf}-\omega _{\ ab}^{b}\widetilde{T}^{afd}-\frac{1}{4}\overline{\psi }(\omega _{i}^{\ fd}\gamma^{i}-\omega _{i}^{\ df}\gamma ^{i})\psi .
\end{equation}
Combining these two terms, Eq. (\ref{DT-t}) can be simplified as:
\begin{eqnarray}
\widetilde{D}_{i}\widetilde{T}^{ifd}+t^{[fd]} &=&\omega _{\ ab}^{f}\widetilde{T}^{abd}-\omega _{\ ab}^{d}\widetilde{T}^{abf}-\omega _{\ ab}^{b}\widetilde{T}^{afd}+\omega _{\ bi}^{i}\widetilde{T}^{bfd}+\Gamma_{\ bi}^{f}\widetilde{T}^{ibd}+\Gamma_{\ bi}^{d}\widetilde{T}^{ifb} 
\nonumber \\
&=&K_{\ ab}^{f}\widetilde{T}^{abd}-K_{\ ab}^{d}\widetilde{T}^{abf}  \nonumber \\
&=&-\frac{1}{2}\left(T_{\ ab}^{f}\widetilde{T}^{abd}-T_{\ ab}^{d}\widetilde{T}^{abf}\right).
\end{eqnarray}
By substituting it into Eq. (\ref{antiequation}), the antisymmetric part  of EOM can be simplified as:

\begin{eqnarray}
\frac{1}{2}\left( \sigma _{fcb}\sigma _{\ \ d}^{cb}-\sigma
_{dcb}\sigma _{\ \ f}^{cb}\right) +\sigma _{a}\left( \sigma_{fd}^{\ \ a}-\sigma _{df}^{\ \ a}\right) &=&\frac{\pi }{2}l_{p}\left(\sigma_{fab}\widetilde{T}^{ab}_{\ \ d}-\sigma_{dab}\widetilde{T}^{ab}_{\ \ f}\right)-\frac{\pi}{2} l_{p}\left( \widetilde{T}_{fab}\Sigma _{d}^{\ ab}-\widetilde{T}_{dab}\Sigma _{f}^{\ ab}\right)\nonumber \\
&=&\frac{\pi }{2}l_{p}\left(\sigma_{fab}\widetilde{T}^{ab}_{\ \ d}-\sigma_{dab}\widetilde{T}^{ab}_{\ \ f}\right)+\frac{\pi}{4} l_{p}\left( \widetilde{T}_{fab}\sigma _{d}^{\ ab}-\widetilde{T}_{dab}\sigma _{f}^{\ ab}\right)\nonumber \\
&=&\frac{\pi}{4}l_{p}\left(\sigma_{fab}\widetilde{T}^{ab}_{\ \ d}-\sigma_{dab}\widetilde{T}^{ab}_{\ \ f}\right). \label{antifinal}
\end{eqnarray}
\bigskip

\end{widetext}

\end{document}